\documentclass[conference]{IEEEtran}
 
\ifCLASSINFOpdf
\usepackage[pdftex]{graphicx}
\else
\usepackage[dvips]{graphicx}
\fi
\usepackage{subcaption}
\usepackage{multirow}
\usepackage{psfrag}

\usepackage[backend=bibtex,doi=false,isbn=false,url=true,bibstyle=ieee,citestyle=numeric-comp,dashed=false]{biblatex}

\usepackage{float}

\usepackage[cmex10]{amsmath}
\usepackage{resizegather}

\usepackage[hidelinks]{hyperref}

\usepackage{cleveref}			
\crefformat{equation}{(#2#1#3)}
\crefrangeformat{equation}{(#3#1#4)--(#5#2#6)}
\crefmultiformat{equation}{(#2#1#3)}%
{,~(#2#1#3)}{, (#2#1#3)}{,~(#2#1#3)}

\usepackage{siunitx}
\DeclareSIUnit \VAr {VAr} 
\DeclareSIUnit \VA {VA} 
\DeclareSIUnit \rad {Radians} 

\usepackage{tikz}			
\usetikzlibrary{shapes.geometric}
\usetikzlibrary{decorations.pathreplacing}		

\usepackage{circuitikz}

\usepackage{url}
\usepackage{breakurl}

\usepackage[]{fancyhdr} %
\newcommand{\changefont}{\fontsize{9}{9}\selectfont}
\allowdisplaybreaks

\IEEEoverridecommandlockouts

\begin{document}

\title{Towards Optimal Integrated Planning of Electricity and Hydrogen Infrastructure for Large-Scale Renewable Energy Transport}

\author{\IEEEauthorblockN{Sleiman~Mhanna,~\emph{Member, IEEE}\\ Isam Saedi,~\emph{Student~Member, IEEE}\\ Guanchi Liu,~\emph{Student~Member, IEEE} \\Pierluigi Mancarella,~\emph{Senior~Member, IEEE}}
\IEEEauthorblockA{Department of Electrical and Electronic Engineering\\The University of Melbourne\\
Melbourne 3010, Australia\\
\{sleiman.mhanna@,isaedi@student.,guanchil@student.,\\pierluigi.mancarella@\}unimelb.edu.au}
}

\maketitle
\thispagestyle{fancy}
\pagestyle{fancy}

\begin{abstract}
	The imminent advent of large-scale green hydrogen (H$_2$) production raises the central question of which of the two options, transporting ``green'' molecules, or transporting ``green'' electrons, is the most cost-effective one. This paper proposes a first-of-its-kind mathematical framework for the optimal integrated planning of electricity and H$_2$ infrastructure for transporting large-scale variable renewable energy (VRE). In contrast to most existing works, this work incorporates essential nonlinearities such as voltage drops due to losses in high-voltage alternating current (HVAC) and high-voltage direct current (HVDC) transmission lines, losses in HVDC converter stations, reactive power flow, pressure drops in pipelines, and linepack, all of which play an important role in determining the optimal infrastructure investment decision. Capturing these nonlinearities requires casting the problem as a nonconvex mixed-integer nonlinear program (MINLP), whose complexity is further exacerbated by its large size due to the relatively high temporal resolution of RES forecasts. This work then leverages recent advancements in convex relaxations to instead solve a tractable alternative in the form of a mixed-integer quadratically constrained programming (MIQCP) problem. The impact of other fundamental factors such as transmission distance and RES capacity is also thoroughly analysed on a canonical two-node system. The integrated planning model is then demonstrated on a real-world case study involving renewable energy zones in Australia.
\end{abstract}

\begin{IEEEkeywords}
	Integrated planning, Hydrogen networks, Linepack, HVDC, HVAC, Renewable energy, MINLP, MIQCP.
\end{IEEEkeywords}

\IEEEpeerreviewmaketitle

\section{Introduction}

Due to the variability of renewable energy sources (RES), maximising their utilisation is arguably one of the biggest challenges facing energy system operators in Australia and around the world. Maximising this utilisation requires energy storage, which, in the case of large energy volumes, may be problematic as large-scale battery storage alone is too costly and pumped-hydro storage is limited for geographical reasons \cite{Gur2018_Reviewofenergystorage}. A promising long-term solution for maximising the integration of VRE consists of building a new infrastructure for \emph{transporting} VRE in the form of electricity and/or H$_2$. Large-scale renewable energy hubs coupled to H$_2$ production hubs may unlock substantial economies of scale predicated on building a \emph{cost-effective} VRE transport infrastructure. Designing a cost-effective infrastructure will need to address the challenging questions of (i) whether VRE hubs and electrolysers should be co-located, (ii) whether to transport VRE as molecules in H$_2$ pipelines or as electrons in electricity transmission lines, and (iii) the drivers and conditions that favour one investment option over another. Answering the above questions is a massive undertaking that requires an integrated electricity and H$_2$ system (IEHS) modelling framework to assess costs and benefits of different investment options. 

As many of the challenges identified here are relatively new, existing knowledge and modelling tools are inadequate for performing such a large-scale \emph{optimal} integrated infrastructure design exercise. In particular, existing state-of-the-art literature is either limited in scope to H$_2$ supply chain only \cite{DeLeonAlmaraz2014_HSC,MorenoBenito2017_HSC,Weber2018_HSCwithhydraulics,Li2019_HSCreview}, i.e., disregarding electricity infrastructure options, or is limited in the variety of considered infrastructure technologies \cite{Samsatli2018_MILPintegratedEHS,Welder2019_OptimizationofEHS,Singlitico2021_H2fromonshoreandoffshore,DeSantis2021_Comparisonoftransportoptions}. Considering all the relevant transport and storage technologies in an integrated framework can unlock superior design solutions. This is especially true when considering the specific features associated with RES, and in particular when they are clustered in large-scale renewable energy hubs where wind and solar farms may be located far from the location of H$_2$ utilisation.

Other essential aspects that are ignored in the literature include voltage drops due to losses in transmission lines, pressure drops in pipelines, linepack,\footnote{The linepack is the amount of pressurised gas stored in a pipeline network.} compressor sizing, water availability for electrolysers, and reactive power compensation, all of which play an important role in determining the optimal infrastructure investment decision. The modelling of the linepack is instrumental in quantifying the VRE storage capacity of the H$_2$ pipeline network, which can in turn influence the sizing of H$_2$ pipelines and compressors. In fact, most (if not all) existing works use \emph{steady-state} gas flow models, which are generally inadequate in gas transmission networks where H$_2$ injections from the VRE introduce time-varying accumulation rates. More importantly, with the exception of \cite{Samsatli2018_MILPintegratedEHS,Welder2019_OptimizationofEHS}, the majority of existing works, including \cite{MorenoBenito2017_HSC,Weber2018_HSCwithhydraulics,Singlitico2021_H2fromonshoreandoffshore,DeSantis2021_Comparisonoftransportoptions}, only examine transport options between just two nodes, as opposed to over a network with a general topology (which may include loops and parallel links).

In light of the knowledge gaps identified above, this paper introduces a novel mathematical optimisation model aiming at finding the optimal integrated infrastructure planning for transporting large-scale VRE as either electricity lines and/or H$_2$ pipelines. Specifically, the model not only considers all relevant infrastructure technologies such as HVDC, HVAC, reactive power plants, and H$_2$ pipelines and compressors, but also incorporates all the essential nonlinearities that directly influence the optimal infrastructure investment decision, such as voltage drops due to losses in HVAC and HVDC transmission lines, losses in HVDC converter stations, reactive power flow, pressure drops in pipelines, and linepack. Additionally, the model adopts a relatively high temporal resolution to fully capture the variability of RES and its impact on the optimal investment decision. Instead of directly solving the resulting large-scale nonconvex mixed-integer nonlinear programming (MINLP) problem, which is computationally intractable, the paper introduces a tractable alternative in the form of a mixed-integer quadratically constrained programming (MIQCP) relaxation. This novel MIQCP model is demonstrated on a set of studies that rigorously analyse the impact of the two fundamental factors, distance and RES capacity, on the optimal planning decision. The MIQCP model is also demonstrated on a real-world case study involving actual renewable energy zones in Australia.

The paper is organised as follows. Section~\ref{sec_modelling} introduces the optimal integrated VRE transport infrastructure design model and Section~\ref{sec_micp} describes how to derive a strong MIQCP relaxation of the problem. Section~\ref{sec_numericaleval} numerically evaluates the proposed MIQCP model on a canonical 2-node system as well as on a real-world case study involving renewable energy zones in Australia. The paper concludes in Section~\ref{sec_conclusion}.

\section{Mathematical modelling}\label{sec_modelling}

\begin{figure}[t]
	\centering{
		\includegraphics[width=\columnwidth] {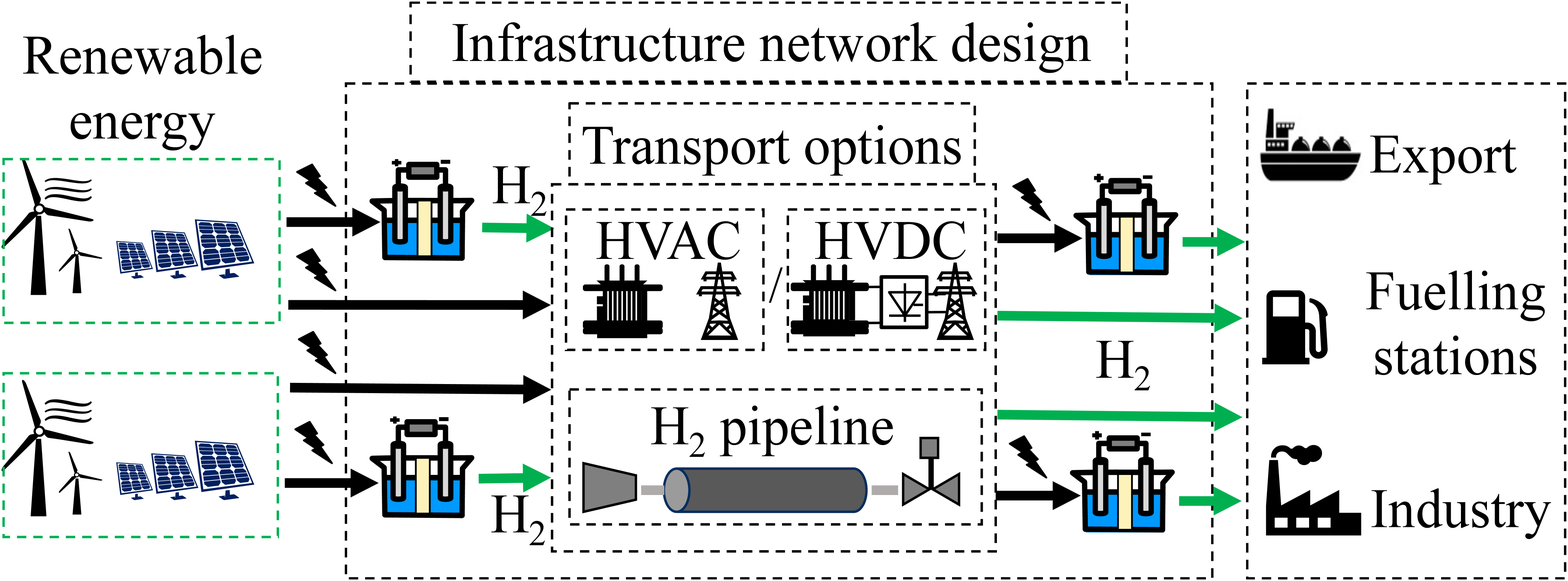}}
	\caption{Illustration of a prototype integrated VRE transport infrastructure design model where the energy from multiple VRE hubs can be transported to an H$_2$ demand point using electricity lines and/or H$_2$ pipelines over a network.}
	\label{fig_VREtransport}
\end{figure}

A prototype integrated VRE transport infrastructure design model is shown in Figure~\ref{fig_VREtransport}, where electricity transmission line options include both HVDC and HVAC, as well as their associated control equipment such as transformers, reactive power compensation, and converters. The H$_2$ pipeline options also include compressors and pressure regulators.

\subsection{Electrolyser station model}\label{sec_ptg}

\begin{figure}[t!]
	\centering{
		\def\pta{0.75}
		\begin{tikzpicture}
			\coordinate (A) at (0,0);
			\coordinate (B) at ({\pta},0);
			\coordinate (C) at ($(B) + (B)$);
			\coordinate (D) at ($(C) + (B)$);
			\coordinate (E) at ($(D) + (B)$);
			\coordinate (F) at ($(E) + (0.5,0)$);
			\coordinate (G) at ($(F) + (B)$);
			
			\filldraw[black] (A) circle (1.5pt) node[above left]{$i$};
			\draw[thick] (A) -- (B);
			\draw[thick] ($(B) - (0,0.4)$) -- ($(B) + (0,0.4)$) -- ($(C) + (0,0.4)$) -- ($(C) - (0,0.4)$) -- cycle;
			\node[] at ($(B) + ({\pta/2},0)$) {PtG};
			\draw[thick] (C) -- (D) -- (E);
			\filldraw[black] (D) circle (1.5pt);
			\draw[->] ($(A) + (0,0.5)$) -- node[above] {$\footnotesize p_{ity}^{\rm ptg}$} ++(B);
			\draw[->] ($(C) + (0,0.5)$) -- node[above] {$\footnotesize \phi_{mty}^{\rm ptg}$} ++(B);
			\draw[->] ($(B) + ({\pta/2},1)$) -- ($(B) + ({\pta/2},0.5)$);
			\node[] at ($(B) + ({\pta/2},1.15)$) {$\footnotesize w_{ity}^{\rm ptg}$};
			\node[below] at (D) {$\footnotesize \wp_{mty}^{\rm ptg}$};
			
			\draw[gray,thick] ($(E) - (0,0.4)$) -- ($(E) + (0,0.4)$) -- ($(F) + (0,0.2)$) -- ($(F) - (0,0.2)$) -- cycle;
			\draw[thick] (F) -- (G);
			\filldraw[black] (G) circle (1.5pt) node[above right]{$m$};
			\node[below] at (G) {$\footnotesize \wp_{mty}^{\vphantom{ptg}}$};
			
			\draw [decorate,decoration={brace,amplitude=5pt,mirror,raise=3.2ex}]
			(A) -- ($(D) - (0.03,0)$) node[midway,yshift=-2.8em]{\small Electrolyser};
			\draw [decorate,decoration={brace,amplitude=5pt,mirror,raise=3.2ex}]
			($(D) + (0.03,0)$) -- (G) node[midway,yshift=-2.8em]{\small Compressor};
	\end{tikzpicture} }
	\caption{Model of an electrolyser station.}
	\label{fig_ptg}
\end{figure}
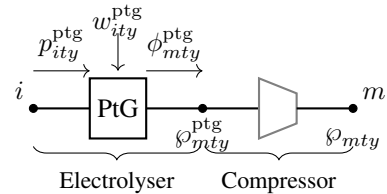

Electrolysers use electricity to split water into H$_2$ and oxygen (O$_2$) in a process called electrolysis. Since the output pressure of a typical proton exchange membrane (PEM) electrolyser is around \SI{3.5}{\mega\pascal} \cite{SiemensEnergyNEB2020}, an electrolyser station in this work is assumed to include a gas compressor to boost the pressure to transmission levels (up to \SI{10}{\mega\pascal}), as shown in Figure~\ref{fig_ptg}.

The decision to install an electrolyser station at a certain location can be captured by a binary variable $z_{im}^{\rm ptg}$ which takes a value of 1 if the electrolyser station is installed and 0 otherwise. In constraint form this can be written as
\begin{align}
	& z_{im}^{\rm ptg} \in \left\lbrace 0,1 \right\rbrace, & im \in \mathcal{E}, \label{eq_electrolyserbinary}
\end{align}
where $\mathcal{E}$ is the set of all candidate electrolyser stations in the network. The process of converting electrical energy to chemical energy can be mathematically written as	
\begin{subequations}\label{eq_powertogas}
	\begin{align}
		& \phi_{mty}^{\rm ptg} = \frac{p_{ity}^{\rm ptg} \eta_{im}^{\rm ptg}}{HHV}, \qquad \qquad im \in \mathcal{E}, \ ty \in \mathcal{T} \times \mathcal{Y} \label{eq_powertovolume} \\
		& w_{ity}^{\rm ptg} = w_{i(t-1)y}^{\rm ptg} - 10\phi_{mty}^{\rm ptg} \rho \Delta \tau, \ im \in \mathcal{E}, \ ty \in \mathcal{T} \times \mathcal{Y} \label{eq_water}
	\end{align}
\end{subequations}
where $p_{ity}^{\rm ptg}$ (\SI{}{\mega\watt}) is the input electrical power to the electrolyser station, $\phi_{mty}^{\rm ptg}$ (\SI{}{\meter\cubed\per\second}) is the aggregated output H$_2$ volumetric flow rate of the electrolyser modules, and $w_{ity}^{\rm ptg}$ (\SI{}{\kilogram}) is the input water consumed by the electrolyser station over a period of $\Delta \tau$ (\SI{}{\second}). In \eqref{eq_powertogas}, $\eta_{im}^{\rm ptg} = 70\%$ is the efficiency of each electrolyser module, $HHV = \SI{12.1948}{\mega\joule\per\meter\cubed}$ is the higher heating value of H$_2$, and $\rho = \SI{0.086}{\kilogram\per\meter\cubed}$ is the density of H$_2$ at standard conditions. The efficiency of the electrolyser station $\eta_{im}^{\rm ptg}$ includes rectifiers and transformers (including transformer cooling and gas cooling). Constraint \eqref{eq_water} is founded on the fact that producing \SI{1}{\kilogram} of H$_2$ requires \SI{10}{\kilogram} of water. Each candidate electrolyser station location in the network is associated with a predetermined initial amount of water $w_{i01}^{\rm ptg}$ (at $t=0$ and $y=1$). The input electrical power is constrained by a maximum predetermined upper limit on the size of the station, $\overline{p}_{i}^{\rm ptg}$, through
\begin{align}
	& 0 \le p_{ity}^{\rm ptg} \le \overline{p}_{i}^{\rm ptg} z_{im}^{\rm ptg}. & im \in \mathcal{E}, \ ty \in \mathcal{T} \times \mathcal{Y} \label{eq_powerlimits}
\end{align}
The size of the compressor can be determined from the required horsepower $p_{mty}^{\rm cp}$ (\SI{}{\mega\watt})
\begin{align}\label{eq_ptgcompressorpower}
	p_{mty}^{{\rm cp}} = \frac{K T Z_{m}^{\rm cp} \gamma \phi_{mty}^{\rm ptg}}{(\gamma - 1)\eta_{im}^{\rm cp}}\left( \left( \frac{\overline{\wp}_{m}}{\wp_{mty}^{\rm ptg}} \right)^{\frac{\gamma - 1}{\gamma}} - 1 \right), \nonumber \\
	im \in \mathcal{E}, \ ty \in \mathcal{T} \times \mathcal{Y} 
\end{align} 
where $\gamma = 1.296$ is the isentropic exponent (dimensionless), $K=0.351121 \times 10^{-3}$ (\SI{}{\mega\joule\per\kelvin\per\meter\cubed}), and $\eta_{im}^{\rm cp}$ (dimensionless) is the overall efficiency of the compressor. The maximum output pressure of the compressor can be set to the maximum operating pressure of the H$_2$ network $\overline{\wp}_{m}= \SI{10}{\mega\pascal}$, which, combined with a fixed output pressure of the electrolyser modules $\wp_{mty}^{\rm ptg} = \SI{3.5}{\mega\pascal}$, makes \eqref{eq_ptgcompressorpower} linear in $p_{mty}^{{\rm cp}}$ and $\phi_{mty}^{\rm ptg}$. Note that at the demand point a compressor is not needed and therefore $p_{mty}^{{\rm cp}} = 0$. Finally, the compressibility factor in \eqref{eq_ptgcompressorpower} is obtained from the Soave-Redlich-Kwong (SRK) equation of state \cite{Soave1972_SRK} with $T = \SI{288.15}{\kelvin}$ as the H$_2$ gas temperature at standard conditions. 

The investment cost of electrolyser stations is given by 
\begin{align*}
	I^{\rm ptg} = \sum_{im \in \mathcal{E}} \left( c_{i,0}^{\rm ptg}z_{im}^{\rm ptg} + c_{i,1}^{\rm ptg} \left\| \left( p_{ity}^{\rm ptg} \right)_{ty \in \mathcal{T} \times \mathcal{Y}} \right\|_{\infty} \right. \\ 
	\left. + c_{m}^{\rm cp} \left\| \left( p_{mty}^{\rm cp} \right)_{ty \in \mathcal{T} \times \mathcal{Y}} \right\|_{\infty} \right),
\end{align*}
where $c_{i,0}^{\rm ptg}$ (\SI{}{\mega\$}) is the base installation cost of an electrolyser station, $c_{i,1}^{\rm ptg}$ (\SI{}{\mega\$\per\mega\watt}) is the unit cost of an electrolyser station, $c_{m}^{\rm cp}$ (\SI{}{\mega\$\per\mega\watt}) is the unit cost of the compressor. The unit cost of an electrolyser station $c_{i,1}^{\rm ptg}$ includes the cost of the step-down transformer and rectifier.

\subsection{H$_2$ pipeline model}\label{sec_pipeline}

\begin{figure}[t!]
	\centering{
		\begin{tikzpicture}
			\coordinate (A) at (0,0);
			\coordinate (B) at (4,0);
			\coordinate (C) at (0,0.2);
			\coordinate (AC) at ($(A) + (C)$);
			\coordinate (BC) at ($(B) + (C)$);
			\coordinate (CC) at ($(C) + (C)$);
			\coordinate (D) at (4,0);
			\coordinate (E) at ($(BC) + (1.5,0)$);
			\coordinate (F) at ($(E) + (0.5,0)$);
			\coordinate (G) at ($(F) + (0.75,0)$);
			
			\draw[thick] (A) -- (B);
			\draw[thick] (CC) -- ($(B) + (CC)$);
			\draw[thick] (C) circle (0.2);
			\draw[thick] (B) arc (-90:90:0.2); 
			\draw[thick, dotted] ($(B) + (CC)$) arc (90:270:0.2);
			\draw[thick] ($(AC) + (-1,0)$) -- (AC);
			\draw[thick,dotted] (BC) -- ($(BC) + (0.2,0)$);
			\filldraw[black] ($(AC) + (-1,0)$) circle (1.5pt) node[above left]{$m$};
			\node[below] at ($(BC) + (1,0)$) {$\footnotesize \wp_{nty}^{o}=\wp_{nty}$};
			\draw[thick] ($(BC) + (0.2,0)$) -- ($(BC) + (1,0)$);
			
			\filldraw[black] ($(BC) + (1,0)$) circle (1.5pt) node[above right]{$n$};
			\draw[->] ($(AC) + (-1,0.5)$) -- node[above] {$\footnotesize \phi_{mnty}^{{\rm in},o}$} ++(1,0);
			\draw[->] ($(BC) + (0,0.5)$) -- node[above] {$\footnotesize \phi_{mnty}^{{\rm out},o}$} ++(1,0);
			\draw[->] ($(AC) + (1.5,0.5)$) -- node[above] {$\footnotesize \phi_{mnty}^{o}$} ++(1,0);
			\node[below] at ($(AC) + (-1,0)$) {$\footnotesize \wp_{mty}=\wp_{mty}^{o}$};
	\end{tikzpicture}}
	\caption{Model of an H$_2$ pipeline link.}
	\label{fig_H2pipeline}
\end{figure}
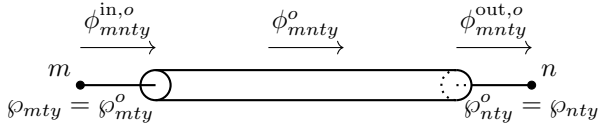

Each gas transmission corridor $mn \in \mathcal{P}$ between junctions $m$ and $n$ is associated with a predetermined set of candidate H$_2$ pipeline link options $\mathcal{O}^{\rm p}$. A model of an H$_2$ pipeline link over gas transmission corridor $mn$ is shown in Figure~\ref{fig_H2pipeline}. 

Different pipeline link options are distinguished by different pipeline diameters including \SI{0.5}{\meter}, \SI{0.9}{\meter}, and \SI{1.2}{\meter}. The gas transmission capacity $\phi_{mnty}^{o}$ (\SI{}{\meter\cubed\per\second}) of a pipeline increases with the diameter $D_{mn}^{o}$ (\SI{}{\meter}). The decision of choosing a certain pipeline option can be captured by a binary variable $z_{mn}^{{\rm p},o}$ which takes a value of 1 if option $o$ is installed and 0 otherwise. In constraint form this can be written as
\begin{align}
	& z_{mn}^{{\rm p},o} \in \left\lbrace 0,1 \right\rbrace, & mn \in \mathcal{P}, \ o \in \mathcal{O}^{\rm p} \label{eq_pipelinebinary}
\end{align}
where $\mathcal{P}$ is the set of all tentative pipeline corridors where H$_2$ gas is flowing from junction $m$ towards junction $n$. In this work, the direction of gas flow is known in advance owing to the predetermined locations of RES and H$_2$ demand (off-take) locations. The average gas volume flow rate $\phi_{mnty}^{o}$ (\SI{}{\meter\cubed\per\second}) across pipeline option $o$ over transmission corridor $mn$ can be obtained from the discretised equation of \emph{motion} along the full length of the pipe \cite{Osiadacz1987_Simulationofgasnetworks}
\begin{align}
	& \left( \phi_{mnty}^{o} \right)^2 = \Phi_{mn}^{o} \left((\wp_{mty}^{o})^2 - (\wp_{nty}^{o})^2 \right), & mn \in \mathcal{P}, \label{eq_motion} 
\end{align}
for all $ o \in \mathcal{O}^{\rm p}$, $ ty \in \mathcal{T} \times \mathcal{Y}$, where
\begin{align*}
	\Phi_{mn}^{o} = \frac{\eta_{mn}^{o}\pi^2 (D_{mn}^{o})^5}{16 \rho^2 Z_{mn}^{o} R T L_{mn}^{o} f_{mn}^{o}},
\end{align*}
and $f_{mn}^{o} = 4\left(20.621 (D_{mn}^{o})^{1/6} \right)^{-2}$ defines the \emph{Weymouth} friction factor \cite{Menon2005_PipelineHydrolics}, and $\eta_{mn}^{o}$ is the pipe efficiency. The compressibility factor $Z_{mn}^{o}$ is computed from the SRK equation of state \cite{Soave1972_SRK}.
In \eqref{eq_motion}, the pressures $\wp_{nty}^{o}$ (\SI{}{\pascal}) are related to the junction pressures $\wp_{nty}$ (\SI{}{\pascal}) through
\begin{align}
	& \underline{\wp}_{m}z_{mn}^{{\rm p},o} \le \wp_{mty}^{o} \le \overline{\wp}_{m}z_{mn}^{{\rm p},o}, &  \label{eq_pressurelimitsofpipem} \\
	& \underline{\wp}_{n}z_{mn}^{{\rm p},o} \le \wp_{nty}^{o} \le \overline{\wp}_{n}z_{mn}^{{\rm p},o}, &  \label{eq_pressurelimitsofpipen} \\
	& \underline{\wp}_{m}\left( 1 - z_{mn}^{{\rm p},o} \right) \le \wp_{mty} - \wp_{mty}^{o} \le \overline{\wp}_{m}\left( 1 - z_{mn}^{{\rm p},o} \right), &  \label{eq_pressurecouplingofpipem} \\
	& \underline{\wp}_{n}\left( 1 - z_{mn}^{{\rm p},o} \right) \le \wp_{nty} - \wp_{nty}^{o} \le \overline{\wp}_{n}\left( 1 - z_{mn}^{{\rm p},o} \right), &  \label{eq_pressurecouplingofpipen} 
\end{align}
for all $mn \in \mathcal{P}$, $ o \in \mathcal{O}^{\rm p}$, $ ty \in \mathcal{T} \times \mathcal{Y}$ and
\begin{align}
	& \underline{\wp}_{m} \le \wp_{mty} \le \overline{\wp}_{m}, & m \in \mathcal{J}, \ ty \in \mathcal{T} \times \mathcal{Y} \label{eq_pressurelimits} 
\end{align}
where $\mathcal{J}$ is the set of all gas junctions in the network. The volumetric gas flows entering and leaving the pipe are related to the average volumetric flow rate across the pipe through 
\begin{align}
	& \phi_{mnty}^{o} = 0.5 \left( \phi_{mnty}^{{\rm in},o} + \phi_{mnty}^{{\rm out},o} \right), & mn \in \mathcal{P} \label{eq_flowinout} \\
	& 0 \le \phi_{mnty}^{o},\phi_{mnty}^{{\rm in},o},\phi_{mnty}^{{\rm out},o} \le \overline{\phi}_{mnty}^{o}, & mn \in \mathcal{P} \label{eq_flowlimits}
\end{align}
for all $ o \in \mathcal{O}^{\rm p}$, $ ty \in \mathcal{T} \times \mathcal{Y}$, and the average pressure across a pipe is defined as
\begin{align}\label{eq_averagepressure}
	\hspace{-0.3cm} \wp_{mnty}^{o} = \frac{2}{3} \left(\wp_{mty}^{o} + \wp_{nty}^{o} - \frac{\wp_{mty}^{o}\wp_{nty}^{o}}{\wp_{mty}^{o} + \wp_{nty}^{o}} \right), \ mn \in \mathcal{P}
\end{align}
for all $ o \in \mathcal{O}^{\rm p}$, $ ty \in \mathcal{T} \times \mathcal{Y}$. The linepack in the pipeline can now be captured by
\begin{subequations}\label{eq_linepack}
	\begin{align}
		& \ell_{mnty}^{o}=\Psi_{mn}^{o} \wp_{mnty}^{o}, & \label{eq_linepack1} \\
		& \ell_{mnty}^{o}=\ell_{mn(t-1)y}^{o} + \Delta \tau \left( \phi_{mnty}^{{\rm in},o} - \phi_{mnty}^{{\rm out},o} \right), & \label{eq_continuity}
	\end{align}
\end{subequations} 
for all $mn \in \mathcal{P}$, $ o \in \mathcal{O}^{\rm p}$, $ ty \in \mathcal{T} \times \mathcal{Y}$, where \eqref{eq_continuity} is the discretised \emph{continuity} equation over the full length $L_{mn}^{o}$ of the pipe and
\begin{align*}
	\Psi_{mn}^{o} = \frac{\pi (D_{mn}^{o})^2 L_{mn}^{o}}{4\rho Z_{mn}^{o} R T}.
\end{align*}
To ensure fairness, the initial value of the linepack (at $t=0$ and $y=1$) for all the pipeline options is set to its minimum value, i.e.,
\begin{align}\label{eq_linepack0}
	& \ell_{mn01}^{o}=\Psi_{mn}^{o} \underline{\wp}_{n} z_{mn}^{{\rm p},o},
\end{align}
for all $mn \in \mathcal{P}$, $ o \in \mathcal{O}^{\rm p}$, $ ty \in \mathcal{T} \times \mathcal{Y}$. Finally, the gas balance equations at each junction of the gas network can now be written as
\begin{multline}\label{eq_gasbalance}
	\left( \phi_{mty}^{\rm ptg} - \mu p_{mty}^{\rm cp} \right) = \\
	\sum_{o \in \mathcal{O}^{\rm p}} \left( \sum_{mn \in \mathcal{P}} \phi_{mnty}^{{\rm in},o} - \sum_{nm \in \mathcal{P}} \phi_{nmty}^{{\rm out},o} \right) + \phi_{mty}^{\rm H_2,d}, 
\end{multline} 
for all $m \in \mathcal{J}$, $ ty \in \mathcal{T} \times \mathcal{Y}$, where $\phi_{mty}^{\rm H_2,d}$ is the H$_2$ volumetric flow rate demand at each junction and $\mu = \SI{0.2624}{\meter\cubed\per\mega\joule}$ is the gas turbine fuel rate coefficient of a centrifugal H$_2$ compressor. The product $\mu p_{mty}^{\rm cp}$ delineates the amount of gas consumed by the compressor in the electrolyser station (see Figure~\ref{fig_ptg}) during the pressure boosting process.

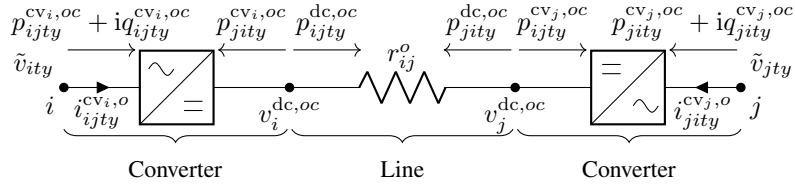
\begin{figure*}[t!]
	\centering{
		\begin{circuitikz}
			\coordinate (A) at (0,0);
			\coordinate (B) at (3,0);
			\coordinate (C) at (6,0);
			\coordinate (D) at (9,0);
			\coordinate (E) at (0.9,0);
			\coordinate (F) at (0.05,0);
			\coordinate (G) at (0,0.5);
			
			\node[below left] at (A) {$i$};
			\node[above left] at (A) {$\footnotesize \tilde{v}_{ity}$};
			\node[below] at (B) {$\footnotesize v_{i}^{{\rm dc},oc}$};
			\node[below] at (C) {$\footnotesize v_{j}^{{\rm dc},oc}$};
			\node[below right] at (D) {$j$};
			\node[above right] at (D) {$\footnotesize \tilde{v}_{jty}$};
			\draw (A) to [sacdc, i>_=$\footnotesize i_{ijty}^{{\rm cv}_{i},o}$, *-*] (B);
			\draw[->] ( $(A) + (F) + (G)$ ) -- node[above] {$\footnotesize p_{ijty}^{{\rm cv}_{i},oc} + \mathrm{i}q_{ijty}^{{\rm cv}_{i},oc}$} ++(E);
			\draw[<-] ( $(B) - (E) - (F) + (G)$ ) -- node[above] {$\footnotesize p_{jity}^{{\rm cv}_{i},oc}$} ++(E);
			\draw (B) to[R=$\footnotesize r_{ij}^{o}$, -*] (C);
			\draw[->] ( $(B) + (F) + (G)$ ) -- node[above] {$\footnotesize p_{ijty}^{{\rm dc},oc}$} ++(E);
			\draw[<-] ( $(C) - (E) - (F) + (G)$ ) -- node[above] {$\footnotesize p_{jity}^{{\rm dc},oc}$} ++(E);
			\draw (C) to [sdcac, i_<=$\footnotesize i_{jity}^{{\rm cv}_{j},o}$, -*] (D);
			\draw[->] ( $(C) + (F) + (G)$ ) -- node[above] {$\footnotesize p_{ijty}^{{\rm cv}_{j},oc}$} ++(E);
			\draw[<-] ( $(D) - (E) - (F) + (G)$ ) -- node[above] {$\footnotesize p_{jity}^{{\rm cv}_{j},oc} + \mathrm{i}q_{jity}^{{\rm cv}_{j},oc}$} ++(E);
			
			\draw [decorate,decoration={brace,amplitude=5pt,mirror,raise=3.5ex}]
			(A) -- ($(B) - (0.03,0)$) node[midway,yshift=-3.1em]{\small Converter};
			\draw [decorate,decoration={brace,amplitude=5pt,mirror,raise=3.5ex}]
			($(B) + (0.03,0)$) -- ($(C) - (0.03,0)$) node[midway,yshift=-3.1em]{\small Line};
			\draw [decorate,decoration={brace,amplitude=5pt,mirror,raise=3.5ex}]
			($(C) + (0.03,0)$) -- (D) node[midway,yshift=-3.1em]{\small Converter};	
	\end{circuitikz}}
	\caption{Model of an HVDC link consisting of two converter stations and a transmission line.}
	\label{fig_HVDCcircuit}
\end{figure*}

The investment cost of the pipeline link is given by 
\begin{align*}
	I^{\rm pipe} = \sum_{mn \in \mathcal{P}} \sum_{o \in \mathcal{O}^{\rm p}} c_{mn}^{{\rm p},o} z_{mn}^{{\rm p},o}, 
\end{align*}
where $c_{mn}^{{\rm p},o}$ (\SI{}{\mega\$}) is the installation cost of a pipeline link of option $o$ over corridor $mn$.

\subsection{HVDC link model}\label{sec_HVDC}

An HVDC link over transmission corridor $ij$ consists of an HVDC transmission line connecting two converter stations, one at the sending end (rectifier) and one at the receiving end (inverter) of the link as shown in Figure~\ref{fig_HVDCcircuit}. Each HVDC transmission corridor $ij \in \mathcal{L}^{\rm dc}_{\rm f}$ between buses $i$ and $j$ is associated with a predetermined set of candidate HVDC link options $\mathcal{O}^{\rm dc}$ and a maximum number of parallel links $c \in \mathcal{C}^{{\rm dc},o} = \left\lbrace 1, \ldots, \overline{c}_{ij}^{{\rm dc},o} \right\rbrace $ for each option $o \in \mathcal{O}^{\rm dc}$. 

Examples of an HVDC transmission link option include a \SI{500}{\kilo\volt} bipole at $1$, $2$, or \SI{3}{\giga\watt} rated capacity. The decision to install a certain option and number of parallel links can be captured by a binary variable $z_{ij}^{{\rm dc},oc}$ which takes a value of 1 if option $o$ and $c$th link are installed and 0 otherwise. It therefore follows that
\begin{align}
	& z_{ij}^{{\rm dc},oc} \in \left\lbrace 0,1 \right\rbrace, & \hspace{-1cm} ij \in \mathcal{L}^{\rm dc}_{\rm f}, \ o \in \mathcal{O}^{\rm dc}, \ c \in \mathcal{C}^{{\rm dc},o} \label{eq_dcbinary} \\
	& \sum_{c \in \mathcal{C}^{{\rm dc},o}} z_{ij}^{{\rm dc},oc} \le \overline{c}_{ij}^{{\rm dc},o}, & ij \in \mathcal{L}^{\rm dc}_{\rm f}, \ o \in \mathcal{O}^{\rm dc} \label{eq_dcsequential1} \\
	& z_{ij}^{{\rm dc},oc} \le z_{ij}^{{\rm dc},o(c-1)}, & ij \in \mathcal{L}^{\rm dc}_{\rm f}, \ c \in \mathcal{C}^{{\rm dc},o} \setminus \{1\} \label{eq_dcsequential2}
\end{align}
where constraints \eqref{eq_dcsequential1} and \eqref{eq_dcsequential2} enforce the sequential installation of links in each option and transmission corridor. In this work, HVDC converter stations are assumed to be of the voltage-source (VSC) type, which can control active and reactive power independently. The main reason for this assumption is that, unlike other converter types such as line commutated converters (LCC), VSC HVDC incorporates self-commutating switching elements that can control active and reactive power independently without additional compensation equipment \cite{Bhattiprolu2021_ACDCTNEP}. This therefore makes VSC HVDC more suitable for transporting renewable energy over long distances, as is the case in this paper. The active, reactive, and apparent power of a VSC are bounded by its ratings and MVA capacity ($\overline{S}^{{\rm cv},o}$) as follows
\begin{subequations}\label{eq_hvdcconverter}
	\begin{align}
		& \underline{p}^{{\rm cv},o}z_{ij}^{{\rm dc},oc} \le p_{ijty}^{{\rm cv}_{i},oc} \le \overline{p}^{{\rm cv},o}z_{ij}^{{\rm dc},oc}, &  \label{eq_converterp} \\
		& \underline{q}^{{\rm cv},o}z_{ij}^{{\rm dc},oc} \le q_{ijty}^{{\rm cv}_{i},oc} \le \overline{q}^{{\rm cv},o}z_{ij}^{{\rm dc},oc}, &  \label{eq_converterq} \\
		& \sqrt{ \left( p_{ijty}^{{\rm cv}_{i},oc} \right)^2 + \left( q_{ijty}^{{\rm cv}_{i},oc} \right)^2 } \le \overline{S}^{{\rm cv},o} z_{ij}^{{\rm dc},oc}, &  
	\end{align}
\end{subequations}
for all $ij \in \mathcal{L}^{\rm dc}_{\rm f} \cup \mathcal{L}^{\rm dc}_{\rm t}$, $ o \in \mathcal{O}^{\rm dc}$, $ c \in \mathcal{C}^{{\rm dc},o}$, $ ty \in \mathcal{T} \times \mathcal{Y}$, where $\mathcal{L}^{\rm dc}_{\rm t}$ is the set of HVDC corridors such that $j$ is the sending-end bus and $i$ is the receiving-end bus. A power electronic converter is an active device whose losses can be obtained from the following parametric equation
\begin{align}
	& p_{ijty}^{{\rm loss}_{i},oc} = \alpha^{{\rm dc},o} z_{ij}^{{\rm dc},oc} + \beta^{{\rm dc},o} i_{ijty}^{{\rm cv}_{i},o} + \gamma^{{\rm dc},o} \left( i_{ijty}^{{\rm cv}_{i},o} \right)^2, \label{eq_converterlosses}
\end{align}
for all $ij \in \mathcal{L}^{\rm dc}_{\rm f} \cup \mathcal{L}^{\rm dc}_{\rm t}$, $ o \in \mathcal{O}^{\rm dc}$, $ c \in \mathcal{C}^{{\rm dc},o}$, $ty \in \mathcal{T} \times \mathcal{Y}$, where $i_{ijty}^{{\rm cv}_{i},o}$ is the magnitude of the AC-side current, $\alpha^{{\rm dc},o}$ captures the no-load losses of transformers and averaged auxiliary equipment losses, $\beta^{{\rm dc},o}$ captures the switching losses of valves and freewheeling diodes as well as some conduction losses due to the series voltage drop, and $\gamma^{{\rm dc},o}$ captures the conduction losses of transformers, switches, and inductors in the VSC. Typical values for the loss parameters are $\alpha^{{\rm dc},o} = \SI{6.62}{\mega\watt}$, $\beta^{{\rm dc},o} = \SI{1800}{\volt}$, and $\gamma^{{\rm dc},o} =\SI{1.98}{\ohm}$ \cite{Daelemans2009_MinLossesusingHVDCVSC}. The converter current $i_{ijty}^{{\rm cv}_{i},o}$ is in turn bounded by
\begin{align}
	& 0 \le i_{ijty}^{{\rm cv}_{i},o} \le \overline{i}^{{\rm cv}_{i},o}z_{ij}^{{\rm dc},oc},&  \label{eq_convertercurrentlimits}
\end{align}
for all $ij \in \mathcal{L}^{\rm dc}_{\rm f} \cup \mathcal{L}^{\rm dc}_{\rm t}$, $o \in \mathcal{O}^{\rm dc}$, $c \in \mathcal{C}^{{\rm dc},o}$, $ty \in \mathcal{T} \times \mathcal{Y}$. The AC and DC sides of the converter are related through
\begin{align}
	& p_{ijty}^{{\rm cv}_{i},oc} + p_{jity}^{{\rm cv}_{i},oc} = p_{ijty}^{{\rm loss}_{i},oc}, & 
	\label{eq_converterACDC}
\end{align}
for all $ij \in \mathcal{L}^{\rm dc}_{\rm f} \cup \mathcal{L}^{\rm dc}_{\rm t}$, $o \in \mathcal{O}^{\rm dc}$, $c \in \mathcal{C}^{{\rm dc},o}$, $ty \in \mathcal{T} \times \mathcal{Y}$, and the DC-side power of the converter is linked to the power flowing through the HVDC transmission line, $p_{ijty}^{{\rm dc},oc}$, through
\begin{align}
	& p_{jity}^{{\rm cv}_{i},oc} + p_{ijty}^{{\rm dc},oc} = 0, & 
	\label{eq_converterDCDC}
\end{align}
for all $ij \in \mathcal{L}^{\rm dc}_{\rm f} \cup \mathcal{L}^{\rm dc}_{\rm t}$, $o \in \mathcal{O}^{\rm dc}$, $c \in \mathcal{C}^{{\rm dc},o}$, $ty \in \mathcal{T} \times \mathcal{Y}$. Additionally, the AC-side current and voltage are related to the active and reactive power injections of the converter through
\begin{align}\label{eq_converterpowerdefinition}
	& \left( p_{ijty}^{{\rm cv}_{i},oc} \right)^2 + \left( q_{ijty}^{{\rm cv}_{i},oc} \right)^2 = \left( v_{ity} \vphantom{i_{ijty}^{{\rm cv}_{i},o}} \right)^2 \left( i_{ijty}^{{\rm cv}_{i},o} \right)^2, & 
\end{align}
for all $ij \in \mathcal{L}^{\rm dc}_{\rm f} \cup \mathcal{L}^{\rm dc}_{\rm t}$, $o \in \mathcal{O}^{\rm dc}$, $c \in \mathcal{C}^{{\rm dc},o}$, $ty \in \mathcal{T} \times \mathcal{Y}$. Finally, the power flowing through the HVDC transmission line, $p_{ijty}^{{\rm dc},oc}$, is defined by
\begin{align}
	& p_{ijty}^{{\rm dc},oc} = \frac{\left( v_{i}^{{\rm dc},oc} \right)^2 - v_{i}^{{\rm dc},oc}v_{j}^{{\rm dc},o}}{r_{ij}^{oc}}, &
	\label{eq_DClinepower}
\end{align}
for all $ij \in \mathcal{L}^{\rm dc}_{\rm f} \cup \mathcal{L}^{\rm dc}_{\rm t}$, $o \in \mathcal{O}^{\rm dc}$, $c \in \mathcal{C}^{{\rm dc},o}$, $ty \in \mathcal{T} \times \mathcal{Y}$, where $r_{ij}^{o}$ is the equivalent resistance of the HVDC bipole transmission line and the DC voltage $v_{i}^{{\rm dc},oc}$ is bounded by
\begin{align}
	& \underline{v}_{i}^{{\rm dc},oc}z_{ij}^{{\rm dc},oc} \le v_{i}^{{\rm dc},oc} \le \overline{v}_{i}^{{\rm dc},oc}z_{ij}^{{\rm dc},oc}, & \label{eq_DClinevoltagelimits}
\end{align}
for all $ij \in \mathcal{L}^{\rm dc}_{\rm f} \cup \mathcal{L}^{\rm dc}_{\rm t}$, $o \in \mathcal{O}^{\rm dc}$, $c \in \mathcal{C}^{{\rm dc},o}$, $ty \in \mathcal{T} \times \mathcal{Y}$. The investment cost of the HVDC transmission link is given by
\begin{align*}
	& I^{\rm HVDC} = \sum_{ij \in \mathcal{L}^{\rm dc}_{\rm f}} \sum_{o \in \mathcal{O}^{\rm dc}} \sum_{c \in \mathcal{C}^{{\rm dc},o}} \left( c_{ij}^{{\rm dc},o} z_{ij}^{{\rm dc},oc} \right), &
\end{align*}	
where $c_{ij}^{{\rm dc},o}$ (\SI{}{\mega\$}) is the investment cost of HVDC link option $o$ over corridor $ij$. The investment cost $c_{ij}^{{\rm dc},o}$ includes the cost of HVDC transmission lines as well as the cost of the two converter stations at the sending-end and receiving-end of the line.

\subsection{HVAC link model}\label{sec_HVAC}

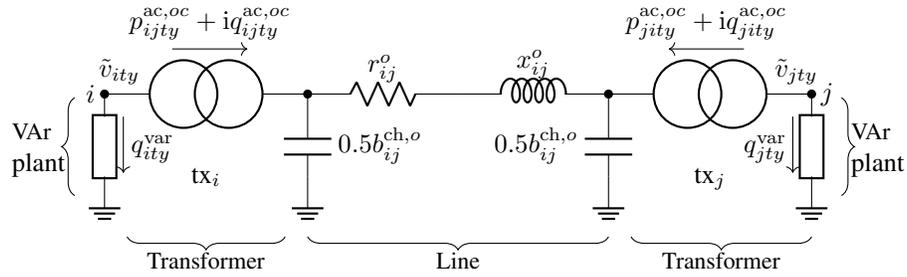
\begin{figure*}[t!]
	\centering{
		\begin{circuitikz}
			\coordinate (A) at (0,0);
			\coordinate (B) at (2.7,0);
			\coordinate (C) at (6,0);
			\coordinate (D) at (9,0);
			\coordinate (E) at (2,0);
			\coordinate (F) at (0.05,0);
			\coordinate (G) at (0,-0.8);
			
			\filldraw[black] (A) circle (1.5pt) node[left]{$i$};
			\filldraw[black] ($(A) + 4*(F)$) node[above]{$\footnotesize \tilde{v}_{ity}$};
			\filldraw[black] ($(B) - 0*(0.15,0)$) circle (1.5pt);
			\draw (A) to[oosourcetrans,/tikz/circuitikz/bipoles/length=2.5cm,l_=tx$_i$] (B);
			\draw[->] ( $(A) + (0.9,0.6)$ ) -- node[above] {$\footnotesize p_{ijty}^{{\rm ac},oc} + \mathrm{i}q_{ijty}^{{\rm ac},oc}$} ++(1,0);
			
			\ctikzset{resistors/scale=0.75}
			\draw (B) to[R, l=$\footnotesize r_{ij}^{o}$] ++(E);
			\draw ($(B) + (E)$) to[L, l=$\footnotesize x_{ij}^{o}$] ($(B) + 2*(E)$);
			\filldraw[black] ($(B) + 2*(E)$) circle (1.5pt);
			\draw ($(B) + 2*(E)$) to[oosourcetrans,/tikz/circuitikz/bipoles/length=2.5cm,l_=tx$_j$] ++(B);
			\filldraw[black] ($2*(B) + 2*(E)$) circle (1.5pt) node[right]{$j$};
			\filldraw[black] ($2*(B) + 2*(E) - 4*(F)$) node[above]{$\footnotesize \tilde{v}_{jty}$};
			\draw[->] ( $2*(B) + 2*(E) + (-0.9,0.6)$ ) -- node[above] {$\footnotesize p_{jity}^{{\rm ac},oc} + \mathrm{i}q_{jity}^{{\rm ac},oc}$} ++(-1,0);
			
			\ctikzset{capacitors/scale=0.75}
			\draw (B) -- ++(0,-0.3) to[C, l=$\footnotesize 0.5b_{ij}^{{\rm ch},o}$] ++(G) node[ground]{};
			\draw ($(B) + 2*(E)$) -- ++(0,-0.3) to[C, l_=$\footnotesize 0.5b_{ij}^{{\rm ch},o}$] ++(G) node[ground]{};
			
			\ctikzset{resistor = european}
			\draw (A) -- ++(0,-0.3) to[R, l^=$ \ \footnotesize q_{ity}^{\rm var}$] ++(G) node[ground]{};
			\draw[->] ($(A) + (0.25,-0.3)$) -- ++(0,-0.75);
			\draw ($2*(B) + 2*(E)$) -- ++(0,-0.3) to[R, l_=$\footnotesize q_{jty}^{\rm var} \ $] ++(G) node[ground]{};
			\draw[->] ($2*(B) + 2*(E) + (-0.25,-0.3)$) -- ++(0,-0.75);
			
			\draw [decorate,decoration={brace,amplitude=5pt,mirror,raise=12ex}]
			(B) -- ($(B) + 2*(E)$) node[midway,yshift=-6.3em]{\small Line};
			\draw [decorate,decoration={brace,amplitude=5pt,mirror,raise=12ex}]
			($(A) + 6*(F)$) -- ($(B) - 6*(F)$) node[midway,yshift=-6.3em]{\small Transformer};
			\draw [decorate,decoration={brace,amplitude=5pt,mirror,raise=12ex}]
			($(B) + 2*(E) + 6*(F)$) -- ++($(B) - 6*(F)$) node[midway,yshift=-6.3em]{\small Transformer};
			
			\draw [decorate,decoration={brace,amplitude=5pt,mirror,raise=2.5ex}]
			($(A) + (0,-0.05)$) -- ++(0,-1.3) node[xshift=-2.5em, label={[align=left]\small VAr\\plant}]{};
			\draw [decorate,decoration={brace,amplitude=5pt,raise=2.5ex}]
			($2*(B) + 2*(E) + (0,-0.05)$) -- ++(0,-1.3) node[xshift=+2.5em, label={[align=left]\small VAr\\plant}]{};
	\end{circuitikz}}
	\caption{Model of an HVAC link consisting of two transformer substations and a transmission line.}
	\label{fig_HVACcircuit}
\end{figure*}

An HVAC link over transmission corridor $ij$ consists of an HVAC transmission line connecting two ideal transformers, one at the sending end and one at the receiving end of the link as shown in Figure~\ref{fig_HVACcircuit}. Each HVAC transmission corridor $ij \in \mathcal{L}^{\rm ac}_{\rm f}$ between buses $i$ and $j$ is associated with a predetermined set of candidate HVAC line options $\mathcal{O}^{\rm ac}$ and a maximum number of parallel links $c \in \mathcal{C}^{{\rm ac},o} = \left\lbrace 1, \ldots, \overline{c}_{ij}^{{\rm ac},o} \right\rbrace $ for each option $o \in \mathcal{O}^{\rm ac}$. 

Examples of an HVAC transmission link option include \SI{345}{\kilo\volt} at \SI{0.75}{\giga\watt} rated capacity, \SI{500}{\kilo\volt} at \SI{1.5}{\giga\watt} rated capacity, and \SI{765}{\kilo\volt} at \SI{1.5}{\giga\watt} rated capacity in both single and double circuit arrangements. The decision of choosing a certain option and number of parallel links (single or double circuits) can be captured by a binary variable $z_{ij}^{{\rm ac},oc}$ which takes a value of 1 if option $o$ and $c$th link are installed and 0 otherwise. It therefore follows that 
\begin{align}
	& z_{ij}^{{\rm ac},oc} \in \left\lbrace 0,1 \right\rbrace, & \hspace{-1cm} ij \in \mathcal{L}^{\rm ac}_{\rm f}, \ o \in \mathcal{O}^{\rm ac}, \ c \in \mathcal{C}^{{\rm ac},o} \label{eq_acbinary} \\
	& \sum_{c \in \mathcal{C}^{{\rm ac},o}} z_{ij}^{{\rm ac},oc} \le \overline{c}_{ij}^{{\rm ac},o}, & ij \in \mathcal{L}^{\rm ac}_{\rm f}, \ o \in \mathcal{O}^{\rm ac} \label{eq_acsequential1} \\
	& z_{ij}^{{\rm ac},oc} \le z_{ij}^{{\rm ac},o(c-1)}, & ij \in \mathcal{L}^{\rm ac}_{\rm f}, \ c \in \mathcal{C}^{{\rm ac},o} \setminus \{1\} \label{eq_acsequential2}
\end{align}
where constraints \eqref{eq_acsequential1} and \eqref{eq_acsequential2} enforce the sequential installation of links in each option and transmission corridor. 

The complex voltage $\tilde{v}_{i}$ (pu) at bus $i$ can be expressed as $\tilde{v}_{i}=v_{i} \mathrm{e}^{\mathrm{i} \theta_{i}}=v_{i} \angle \theta_{i} = v_{i}\cos\left(\theta_{i}\right) + \mathrm{i} v_{i}\sin\left(\theta_{i}\right)$ in \emph{polar form}, where $\mathrm{i} = \sqrt{-1}$. HVAC transmission lines and phase-shifting transformers are represented by their $\pi$-model equivalents, in which the admittance is defined as $\tilde{y}_{ij}=1/(r_{ij} + \mathrm{i}x_{ij}) = g_{ij} + \mathrm{i} b_{ij}$, where $g_{ij}$ and $b_{ij}$ are the conductance (pu) and susceptance (pu), respectively. Additionally, the charging susceptance in the $\pi$-model of branch $ij$ is denoted by $b^{\rm ch}_{ij}$ (pu). By defining
\begin{align*}
	w_{i} & =\left|\tilde{v}_{i}\right|^2 = v^{2}_{i}, & i \in \mathcal{B} \\
	w_{ij}^{\rm r} & =\Re\left\{\tilde{v}_{i} \tilde{v}_{j}^*\right\}=v_{i}v_{j}\cos\left(\theta_{i}-\theta_{j}\right), & ij \in \mathcal{L} \\
	w_{ij}^{\rm i} & =\Im\left\{\tilde{v}_{i} \tilde{v}_{j}^*\right\}=v_{i}v_{j}\sin\left(\theta_{i}-\theta_{j}\right), & ij \in \mathcal{L}
\end{align*}
the active and reactive power flows over link $c$ of option $o$ in corridor $ij$ can be written as
\begin{subequations}\label{eq_pijqij}	
	\begin{align}
		& p_{ijty}^{{\rm ac},oc} = g^{{\rm c},o}_{ij} w_{ity}^{{\rm ac},oc} - g_{ij}^{o} w_{ijty}^{{\rm r},oc} + b_{ij}^{o} w_{ijty}^{{\rm r},oc}, & \label{eq_pij} \\
		& q_{ijty}^{{\rm ac},oc} = b^{{\rm c},o}_{ij} w_{ity}^{{\rm ac},oc} - b_{ij}^{o} w_{ijty}^{{\rm r},oc} - g_{ij}^{o} w_{ijty}^{{\rm r},oc}, & \label{eq_qij} 
	\end{align}
\end{subequations} 
for all $ij \in \mathcal{L}^{\rm ac}_{\rm f} \cup \mathcal{L}^{\rm ac}_{\rm t}$, $oc \in \mathcal{O}^{\rm ac} \times \mathcal{C}^{{\rm ac},o}$, $ty \in \mathcal{T} \times \mathcal{Y}$, where $g^{{\rm c},o}_{ij}=\Re\{ (\tilde{y}_{ij}^{o})^*-0.5\mathrm{i}b^{{\rm ch},o}_{ij} \} = g_{ij}^{o}$, $b^{{\rm c},o}_{ij}=\Im \{ (\tilde{y}_{ij}^{o})^*-0.5\mathrm{i}b^{{\rm ch},o}_{ij} \} = -b_{ij}^{o} - 0.5b^{{\rm ch},o}_{ij}$, and $\mathcal{L}^{\rm ac}_{\rm t}$ is the set of HVAC corridors such that $j$ is the sending-end bus and $i$ is the receiving-end bus.\footnote{Note that $w_{ijty}^{{\rm r},oc} = -w_{jity}^{{\rm r},oc}$.} In \eqref{eq_pijqij}, $w_{ity}^{{\rm ac},oc}$, $w_{ijty}^{{\rm r},oc}$, and $w_{ijty}^{{\rm i},oc}$ are set to zero if link $c$ of option $o$ in corridor $ij$ is not installed through
\begin{subequations}\label{eq_wijzij}
	\begin{align}
		& \underline{v}_{i}^{2}z_{ij}^{{\rm ac},oc} \le w_{ity}^{{\rm ac},oc} \le \overline{v}_{i}^{2} z_{ij}^{{\rm ac},oc}, & \\
		& \underline{v}_{j}^{2}z_{ij}^{{\rm ac},oc} \le w_{jty}^{{\rm ac},oc} \le \overline{v}_{j}^{2} z_{ij}^{{\rm ac},oc}, &
	\end{align}
\end{subequations}
for all $ij \in \mathcal{L}^{\rm ac}_{\rm f}$, $oc \in \mathcal{O}^{\rm ac} \times \mathcal{C}^{{\rm ac},o}$, $ty \in \mathcal{T} \times \mathcal{Y}$, and
\begin{subequations}\label{eq_wrwi}
	\begin{align}
		& \underline{v}_{i}\underline{v}_{j}\cos \left( \overline{\theta}_{ij} \right)z_{ij}^{{\rm ac},oc} \le w_{ijty}^{{\rm r},oc} \le \overline{v}_{i}\overline{v}_{j}z_{ij}^{{\rm ac},oc}, & \label{eq_wrzij} \\ 
		& \overline{v}_{i}\overline{v}_{j}\sin \left( \underline{\theta}_{ij} \right)z_{ij}^{{\rm ac},oc} \le w_{ijty}^{{\rm i},oc} \le \overline{v}_{i}\overline{v}_{j}\sin \left( \overline{\theta}_{ij} \right)z_{ij}^{{\rm ac},oc}, &  \label{eq_wizij} 
	\end{align}
\end{subequations}
for all $ij \in \mathcal{L}^{\rm ac}_{\rm f}$, $oc \in \mathcal{O}^{\rm ac} \times \mathcal{C}^{{\rm ac},o}$, $ty \in \mathcal{T} \times \mathcal{Y}$. Conversely, if link $c$ of option $o$ in corridor $ij$ is installed, $w_{ity}^{{\rm ac},oc}$ is set equal to $w_{ity}$ (which is another design variable) through
\begin{subequations}\label{eq_wijwij}
	\begin{align}
		& \underline{v}_{i}^{2} \left( 1 - z_{ij}^{{\rm ac},oc} \right) \le w_{ity} - w_{ity}^{{\rm ac},oc} \le \overline{v}_{i}^{2} \left( 1 - z_{ij}^{{\rm ac},oc} \right), & \\
		& \underline{v}_{j}^{2} \left( 1 - z_{ij}^{{\rm ac},oc} \right) \le w_{jty} - w_{jty}^{{\rm ac},oc} \le \overline{v}_{j}^{2} \left( 1 - z_{ij}^{{\rm ac},oc} \right),
	\end{align}
\end{subequations}
for all $ij \in \mathcal{L}^{\rm ac}_{\rm f}$, $oc \in \mathcal{O}^{\rm ac} \times \mathcal{C}^{{\rm ac},o}$, $ty \in \mathcal{T} \times \mathcal{Y}$, and
\begin{align}
	& \underline{v}_{i}^{2} \le w_{ity} \le \overline{v}_{i}^{2}. & i \in \mathcal{B}, \ ty \in \mathcal{T} \times \mathcal{Y} \label{eq_voltagelimits}
\end{align}
Constraints \cref{eq_wijzij,eq_wrwi,eq_wijwij} ensure that $p_{ijty}^{{\rm ac},oc}$ and $q_{ijty}^{{\rm ac},oc}$ in \eqref{eq_pijqij} are set to zero when the $c$th link of option $o$ in corridor $ij$ is not installed, by setting the corresponding $z_{ij}^{{\rm ac},oc} = 0$ and therefore $w_{ity}^{{\rm ac},oc}$, $w_{ijty}^{{\rm r},oc}$, and $w_{ijty}^{{\rm i},oc}$ all to zero. 

Additionally, links installed in parallel in option $o$ along corridor $ij$ should have their $w_{ijty}^{{\rm r},oc}$ and $w_{ijty}^{{\rm i},oc}$ equal to $w_{ijty}^{{\rm r},o1}$ and $w_{ijty}^{{\rm i},o1}$, respectively, as follows
\begin{subequations}\label{eq_wrc12}
	\begin{align}
		& 0 \le w_{ijty}^{{\rm r},o1} - w_{ijty}^{{\rm r},oc} \le \overline{v}_{i}\overline{v}_{j} \left( 1 - z_{ij}^{{\rm ac},oc} \right), & \label{eq_wrc1} \\
		& \overline{v}_{i}\overline{v}_{j}\sin \left( \underline{\theta}_{ij} \right) \left( 1 - z_{ij}^{{\rm ac},oc} \right) \le w_{ijty}^{{\rm i},o1} - w_{ijty}^{{\rm i},oc}, & \label{eq_wic1} \\
		& w_{ijty}^{{\rm i},o1} - w_{ijty}^{{\rm i},oc} \le \overline{v}_{i}\overline{v}_{j}\sin \left( \overline{\theta}_{ij} \right) \left( 1 - z_{ij}^{{\rm ac},oc} \right), & \label{eq_wic2} 
	\end{align}
\end{subequations}
for all $ij \in \mathcal{L}^{\rm ac}_{\rm f}$, $o \in \mathcal{O}^{\rm ac}$, $c \in \mathcal{C}^{{\rm ac},o} \setminus \{1\}$, $ty \in \mathcal{T} \times \mathcal{Y}$. Since all installed parallel links are constrained by \cref{eq_wijwij} and \cref{eq_wrc12} to all have the same values for $w_{ity}^{{\rm ac},oc}$, $w_{ijty}^{{\rm r},oc}$, and $w_{ijty}^{{\rm i},oc}$ as $w_{ity}^{o1}$, $w_{ijty}^{{\rm r},o1}$ and $w_{ijty}^{{\rm i},o1}$, respectively, it suffices to enforce the (nonconvex) rotated second-order cone constraints 
\begin{align}
	& w_{ity}^{o1}w_{jty}^{o1} = \left( w_{ijty}^{{\rm r},o1} \vphantom{w_{ijty}^{{\rm i},o1}} \right)^2 + \left( w_{ijty}^{{\rm i},o1} \right)^2 , & \label{eq_rsoc1} 
\end{align}
for all $ij \in \mathcal{L}^{\rm ac}_{\rm f}$, $o \in \mathcal{O}^{\rm ac}$, $ty \in \mathcal{T} \times \mathcal{Y}$, the angle difference constraints
\begin{align}
	& \tan \left( \underline{\theta}_{ij} \right) w_{ijty}^{{\rm r},o1} \le w_{ijty}^{{\rm i},o1} \le \tan \left( \overline{\theta}_{ij} \right) w_{ijty}^{{\rm r},o1}, & \label{eq_anglediff1}
\end{align}
for all $ij \in \mathcal{L}^{\rm ac}_{\rm f}$, $o \in \mathcal{O}^{\rm ac}$, $ty \in \mathcal{T} \times \mathcal{Y}$, and the apparent power limit constraints
\begin{align}
	& \sqrt{\left( p_{ijty}^{{\rm ac},o1}\right)^2 + \left( q_{ijty}^{{\rm ac},o1} \right)^2 } \le \overline{S}^{{{\rm ac},o1}}z_{ij}^{{\rm ac},o1} & \label{eq_MVA1} 
\end{align}
for all $ij \in \mathcal{L}^{\rm ac}_{\rm f} \cup \mathcal{L}^{\rm ac}_{\rm t}$, $o \in \mathcal{O}^{\rm ac}$, $ty \in \mathcal{T} \times \mathcal{Y}$, on the first circuit only. The angle difference constraints in \eqref{eq_anglediff1} are necessary to ensure angular displacement is below \SI{45}{\degree} to maintain transient stability of the HVAC power system.
As identified in \cite{Jabr2013_TNEPoptimization}, enforcing constraints \cref{eq_rsoc1,eq_anglediff1,eq_MVA1} on all links can further improve the lower bound at the root node of a MIQCP branch-and-bound algorithm and the effect on computing time can be favorable only if the subsequent reduction in the number of solved continuous problems at each node of the branch-and-bound tree is \emph{not} offset by the increase in their size. Note that since there are no cycles (loops) in this expansion planning setting the (nonconvex) cycle constraints would be redundant and can therefore be ignored without affecting the feasibility of the solution (refer to \cite{Mhanna2021_SLPforOPF} for more detail). 

The nodal active power balance constraints can now be written as
\begin{align}
	& p^{\rm res}_{ity} = \sum_{j \in \mathcal{B}_{i}} \left( \sum_{oc \in \mathcal{O}^{\rm ac} \times \mathcal{C}^{{\rm ac},o}} p_{ijty}^{{\rm ac},oc} \right. + \nonumber \\
	& \left. \sum_{oc \in \mathcal{O}^{\rm dc} \times \mathcal{C}^{{\rm dc},o}} p_{ijty}^{{\rm cv}_{i},oc} \right) + p_{ity}^{\rm ptg}, \ i \in \mathcal{B}, \ ty \in \mathcal{T} \times \mathcal{Y} \label{eq_pbalance} 
\end{align}
where $\mathcal{B}_{i}$ is the set of HVAC buses adjacent to bus $i$, and $p^{\rm res}_{ity}$ is the power injection from wind and solar assets connected to bus $i$ during time slot $t$. This RES power injection is in turn bounded by 
\begin{align}\label{eq_RESbounds}
	0 \le p^{\rm res}_{ity} \le \overline{p}^{\rm res}_{ity}, \ i \in \mathcal{B}, \qquad ty \in \mathcal{T} \times \mathcal{Y}
\end{align}
where $\overline{p}^{\rm res}_{ity}$ is the total generated VRE (wind and solar) at bus $i$ during time slot $t$.

The HVAC planning options typically include reactive power compensation at some HVAC buses, which can improve line loadability by maintaining voltages near rated values and angular displacement below \SI{45}{\degree}. A shunt reactor, or more generally a static VAr compensator (SVC), at bus $i$ can be modelled as
\begin{align}
	& \underline{q}_{i}^{\rm var} z_{i}^{\rm var} \le q_{ity}^{\rm var} \le \overline{q}_{i}^{\rm var} z_{i}^{\rm var}, \ i \in \mathcal{B}, & \ ty \in \mathcal{T} \times \mathcal{Y} \label{eq_qvar} 
\end{align}
where $q_{ity}^{\rm var}$ denotes the reactive power output of a VAr plant at bus $i$ at time $ty \in \mathcal{T} \times \mathcal{Y}$ and $z_{i}^{\rm var}$ is a binary variable that takes a value of 1 when the shunt reactor is installed and 0 otherwise. The nodal reactive power balance constraints can now be written as
\begin{align}
	& - q_{ity}^{\rm var}= \sum_{j \in \mathcal{B}_{i}} \left( \sum_{oc \in \mathcal{O}^{\rm ac} \times \mathcal{C}^{{\rm ac},o}} q_{ijty}^{{\rm ac},oc} \right. + \nonumber \\
	&\left. \sum_{oc \in \mathcal{O}^{\rm dc} \times \mathcal{C}^{{\rm dc},o}} q_{ijty}^{{\rm cv}_{i},oc} \right). \ i \in \mathcal{B}, \ ty \in \mathcal{T} \times \mathcal{Y} \label{eq_qbalance} 
\end{align}
Finally, the investment cost of the combined HVAC transmission links and VAr plants is given by
\begin{multline*}
	I^{\rm HVAC} = \sum_{ij \in \mathcal{L}^{\rm ac}_{\rm f}} \sum_{o \in \mathcal{O}^{\rm ac}} \sum_{c \in \mathcal{C}^{{\rm ac},o}} \left( c_{ij}^{{\rm ac},o} z_{ij}^{{\rm ac},oc} \right) + \\
	\sum_{i \in \mathcal{B}} \left( c_{i,0}^{\rm var} z_{i}^{\rm var} + c_{i,1}^{\rm var} \left\| \left( q_{ity}^{\rm var} \right)_{ty \in \mathcal{T} \times \mathcal{Y}} \right\|_{\infty} \right),
\end{multline*}	
where $c_{ij}^{{\rm ac},o}$ (\SI{}{\mega\$}) is the investment cost of HVAC link option $o$ over corridor $ij$, $c_{i,0}^{\rm var}$ (\SI{}{\mega\$}) is the installation cost the VAr device (SVC) at bus $i$, and $c_{i,1}^{\rm var}$ (\SI{}{\mega\$\per\mega\VAr}) is the unit cost of reactive power output from the VAr device at bus $i$. The investment cost $c_{ij}^{{\rm ac},o}$ includes the cost of HVAC transmission lines as well as the cost of step-up and step-down transformer substations tx$_{i}$ and tx$_{j}$ at the sending-end and receiving-end of the line.

\subsection{Optimal integrated infrastructure planning}\label{sec_optimalplanning}

Mathematically, the objective of the integrated infrastructure planning problem is to simultaneously minimise the total investment cost and maximise the H$_2$ sale over the whole planning horizon $ \mathcal{Y} = \{ 1, \ldots , Y \} $ as
\begin{subequations}\label{OptimalIntegratedPlanning}
	\begin{align}
		& \underset{\substack{x}} 
		{\mbox{minimise}} \quad I^{\rm ptg} + I^{\rm pipe} + I^{\rm HVDC} + I^{\rm HVAC} - \nonumber \\
		& \qquad \qquad \quad \sum_{y \in \mathcal{Y}} \sum_{t \in \mathcal{T}} \sum_{m \in \mathcal{J}} \frac{c_{m}^{\rm H_2} \phi_{mty}^{\rm H_2,d} \Delta \tau}{(1+\iota)^{y}} \label{eq_objective} \\
		& \text{subject to \cref{eq_electrolyserbinary,eq_powertogas,eq_powerlimits,eq_ptgcompressorpower,eq_pipelinebinary,eq_motion,eq_pressurelimitsofpipem,eq_pressurelimitsofpipen,eq_pressurecouplingofpipem,eq_pressurecouplingofpipen,eq_pressurelimits,eq_flowinout,eq_flowlimits,eq_averagepressure,eq_linepack,eq_linepack0,eq_gasbalance,eq_dcbinary,eq_dcsequential1,eq_dcsequential2,eq_hvdcconverter,eq_converterlosses,eq_convertercurrentlimits,eq_converterACDC,eq_converterDCDC,eq_converterpowerdefinition,eq_DClinepower,eq_DClinevoltagelimits,eq_acbinary,eq_acsequential1,eq_acsequential2,eq_pijqij,eq_wijzij,eq_wrwi,eq_wijwij,eq_voltagelimits,eq_wrc12,eq_rsoc1,eq_anglediff1,eq_MVA1,eq_RESbounds,eq_pbalance,eq_qvar,eq_qbalance}},
	\end{align}
\end{subequations}
where $c_{m}^{\rm H_2}$ (\SI{}{\$\per\meter\cubed}) is the selling price (profit) of H$_2$, $\iota$ is the discount rate, and $x$ is a vector that concatenates all the variables of the problem.\footnote{The original selling price is typically in \SI{}{\$\per\kilogram} but is then converted to \SI{}{\$\per\meter\cubed} for consistency of dimensions.}


\section{Mixed-integer convex relaxation}\label{sec_micp} 

Due to the nonconvex nonlinear constraints in \eqref{eq_motion}, \eqref{eq_averagepressure}, \eqref{eq_converterlosses}, \eqref{eq_converterpowerdefinition}, \eqref{eq_DClinepower} and \eqref{eq_rsoc1}, Problem~\ref{OptimalIntegratedPlanning} belongs to the class of mixed-integer nonlinear programming (MINLP) problems that have a nonconvex continuous relaxation, thus making it extremely difficult to solve to global, or even local, optimality. To make matters worse, Problem~\ref{OptimalIntegratedPlanning} requires disjunctive constraints (to encode ``or'' statements) associated with different design variables that correspond to the optimal choice of transport option. These disjunctive constraints require a Big-M reformulation to transform them into MILP constraints such as the ones in \cref{eq_pressurelimitsofpipem,eq_pressurelimitsofpipen,eq_pressurecouplingofpipem,eq_pressurecouplingofpipen}, \cref{eq_wijzij}, \cref{eq_wijwij}, and \cref{eq_wrc12}. Unfortunately, Big-M reformulations are notorious for having weak root node relaxations in general. As a result, this class of MINLPs in particular is intractable even for small scale problems.

Luckily, there exists a tractable alternative to Problem~\ref{OptimalIntegratedPlanning} in the form of a strong mixed-integer quadratically constrained programming (MIQCP) problem. It is straightforward to check that the nonconvex second-order cone (SOC) constraint in \eqref{eq_motion} becomes convex when it is relaxed into an inequality constraint of the form
\begin{align}
	& \left( \phi_{mnty}^{o} \right)^2 \le \Phi_{mn}^{o} \left((\wp_{mty}^{o})^2 - (\wp_{nty}^{o})^2 \right), & mn \in \mathcal{P} \label{eq_motion_r} 
\end{align}
for all $ o \in \mathcal{O}^{\rm p}$, $ ty \in \mathcal{T} \times \mathcal{Y}$. Moreover, because the nodal pressures are nonnegative, constraint \cref{eq_averagepressure} becomes convex when it is relaxed into an inequality constraint of the form
\begin{align}\label{eq_averagepressure_r}
	\hspace{-0.25cm} \wp_{mnty}^{o} \ge \frac{2}{3} \left(\wp_{mty}^{o} + \wp_{nty}^{o} - \frac{\wp_{mty}^{o}\wp_{nty}^{o}}{\wp_{mty}^{o} + \wp_{nty}^{o}} \right), \ mn \in \mathcal{P}
\end{align}
for all $ o \in \mathcal{O}^{\rm p}$, $ ty \in \mathcal{T} \times \mathcal{Y}$.\footnote{The proof can be found in \cite{Mhanna2021_SLPforIEGS}.} However, although convex, constraint \eqref{eq_averagepressure_r} cannot be directly handled by state-of-the-art MIQCP solvers such as \textsc{Gurobi} \cite{Gurobi2019}. For this reason, constraint \eqref{eq_averagepressure_r} can instead be replaced by a tight polyhedral envelope 
\begin{align}\label{eq_averagepressure_r_poly}
	\hspace{-0.25cm} \wp_{mnty}^{o} \ge \frac{2}{3} \left(\wp_{mty}^{o} + \wp_{nty}^{o} + {\rm conv} \left( - \frac{\wp_{mty}^{o}\wp_{nty}^{o}}{\wp_{mty}^{o} + \wp_{nty}^{o}} \right) \right) 
\end{align}
for all $mn \in \mathcal{P}$, $ o \in \mathcal{O}^{\rm p}$, $ ty \in \mathcal{T} \times \mathcal{Y}$, as detailed in \cite{Mhanna2021_SLPforIEGS}.

The first step towards conferring a strong convex relaxation property to constraints \eqref{eq_converterlosses}, \eqref{eq_converterpowerdefinition}, and \eqref{eq_DClinepower} is to define new variables to substitute the square of the voltage and the square of the current terms, i.e., $w := v^2$ and $l:=i^2$, respectively. Constraints \eqref{eq_converterlosses}, \eqref{eq_converterpowerdefinition}, and \eqref{eq_DClinepower} can now be equivalently rewritten as
\begin{align}
	& p_{ijty}^{{\rm loss}_{i},oc} = \alpha^{{\rm dc},o} z_{ij}^{{\rm dc},oc} + \beta^{{\rm dc},o} i_{ijty}^{{\rm cv}_{i},o} + \gamma^{{\rm dc},o} l_{ijty}^{{\rm cv}_{i},o}, \label{eq_converterlosses_li} \\
	& l_{ijty}^{{\rm cv}_{i},o} = \left( i_{ijty}^{{\rm cv}_{i},o} \right)^2 , \label{eq_isquared} \\
	& \left( p_{ijty}^{{\rm cv}_{i},oc} \right)^2 + \left( q_{ijty}^{{\rm cv}_{i},oc} \right)^2 = w_{ity} l_{ijty}^{{\rm cv}_{i},o} , & \label{eq_converterpowerdefinition_wi} \\
	& p_{ijty}^{{\rm dc},oc} = \frac{ w_{i}^{{\rm dc},oc} - w_{ij}^{{\rm dc},o}}{r_{ij}^{oc}}, &
	\label{eq_DClinepower_ww}
\end{align}
for all $ij \in \mathcal{L}^{\rm dc}_{\rm f} \cup \mathcal{L}^{\rm dc}_{\rm t}$, $o \in \mathcal{O}^{\rm dc}$, $c \in \mathcal{C}^{{\rm dc},o}$, $ty \in \mathcal{T} \times \mathcal{Y}$, and
\begin{align}
	&  w_{i}^{{\rm dc},oc} w_{j}^{{\rm dc},oc} = \left( w_{ij}^{{\rm dc},o} \right)^2, &
	\label{eq_wiwj_wij}
\end{align}
for all $ij \in \mathcal{L}^{\rm dc}_{\rm f}$, $o \in \mathcal{O}^{\rm dc}$, $c \in \mathcal{C}^{{\rm dc},o}$, $ty \in \mathcal{T} \times \mathcal{Y}$. It is now straightforward to verify that the nonconvex quadratic constraint in \eqref{eq_isquared} and the nonconvex rotated SOC constraints in \eqref{eq_converterpowerdefinition_wi}, and \eqref{eq_wiwj_wij} become convex when they are relaxed into inequality constraints of the form
\begin{align}
	& l_{ijty}^{{\rm cv}_{i},o} \ge \left( i_{ijty}^{{\rm cv}_{i},o} \right)^2 , \label{eq_isquared_r} \\
	& \left( p_{ijty}^{{\rm cv}_{i},oc} \right)^2 + \left( q_{ijty}^{{\rm cv}_{i},oc} \right)^2 \le w_{ity} l_{ijty}^{{\rm cv}_{i},o} , & \label{eq_converterpowerdefinition_wi_r} 
\end{align}
for all $ij \in \mathcal{L}^{\rm dc}_{\rm f} \cup \mathcal{L}^{\rm dc}_{\rm t}$, $o \in \mathcal{O}^{\rm dc}$, $c \in \mathcal{C}^{{\rm dc},o}$, $ty \in \mathcal{T} \times \mathcal{Y}$ and
\begin{align}
	&  w_{i}^{{\rm dc},oc} w_{j}^{{\rm dc},oc} \ge \left( w_{ij}^{{\rm dc},o} \right)^2, &
	\label{eq_wiwj_wij_r}
\end{align}
for all $ij \in \mathcal{L}^{\rm dc}_{\rm f}$, $o \in \mathcal{O}^{\rm dc}$, $c \in \mathcal{C}^{{\rm dc},o}$, $ty \in \mathcal{T} \times \mathcal{Y}$. Finally, the nonconvex rotated SOC constraint \eqref{eq_rsoc1} can also be similarly relaxed into a convex constraint of the form  
\begin{align}
	& w_{ity}^{o1}w_{jty}^{o1} \ge \left( w_{ijty}^{{\rm r},o1} \vphantom{w_{ijty}^{{\rm i},o1}} \right)^2 + \left( w_{ijty}^{{\rm i},o1} \right)^2 , & \label{eq_rsoc1_r} 
\end{align}
for all $ij \in \mathcal{L}^{\rm ac}_{\rm f}$, $o \in \mathcal{O}^{\rm ac}$, $ty \in \mathcal{T} \times \mathcal{Y}$. The MIQCP relaxation of Problem~\ref{OptimalIntegratedPlanning} can now be written as
\begin{subequations}\label{OptimalIntegratedPlanning_MIQCP}
	\begin{align}
		& \underset{\substack{x}} 
		{\mbox{minimise}} \quad I^{\rm ptg} + I^{\rm pipe} + I^{\rm HVDC} + I^{\rm HVAC} - \nonumber \\
		& \qquad \qquad \quad \sum_{y \in \mathcal{Y}} \sum_{t \in \mathcal{T}} \sum_{m \in \mathcal{J}} \frac{c_{m}^{\rm H_2} \phi_{mty}^{\rm H_2,d} \Delta \tau}{(1+\iota)^{y}} \label{eq_objective_miqcp} \\
		& \text{subject to \cref{eq_electrolyserbinary,eq_powertogas,eq_powerlimits,eq_ptgcompressorpower,eq_pipelinebinary},\cref{eq_pressurelimitsofpipem,eq_pressurelimitsofpipen,eq_pressurecouplingofpipem,eq_pressurecouplingofpipen,eq_pressurelimits,eq_flowinout,eq_flowlimits},\cref{eq_linepack,eq_linepack0,eq_gasbalance,eq_dcbinary,eq_dcsequential1,eq_dcsequential2,eq_hvdcconverter},\cref{eq_convertercurrentlimits,eq_converterACDC,eq_converterDCDC}}, \nonumber \\
		& \text{\cref{eq_DClinevoltagelimits,eq_acbinary,eq_acsequential1,eq_acsequential2,eq_pijqij,eq_wijzij,eq_wrwi,eq_wijwij,eq_voltagelimits,eq_wrc12},\cref{eq_anglediff1,eq_MVA1,eq_RESbounds,eq_pbalance,eq_qvar,eq_qbalance},\cref{eq_motion_r},\cref{eq_averagepressure_r_poly},\cref{eq_converterlosses_li},\cref{eq_DClinepower_ww},\cref{eq_isquared_r,eq_converterpowerdefinition_wi_r,eq_wiwj_wij_r,eq_rsoc1_r}}.
	\end{align}
\end{subequations}

Although Problem~\ref{OptimalIntegratedPlanning_MIQCP} is a relaxation of Problem~\ref{OptimalIntegratedPlanning}, which entails that its solution will most likely be infeasible in the original space of Problem~\ref{OptimalIntegratedPlanning}, \textcolor{black}{this relaxation is expected to provide a high-quality lower bound on the globally optimal solution of Problem~\ref{OptimalIntegratedPlanning}. This is especially true when the solution of this MIQCP relaxation is compared to solutions from MILP approximations \cite{Samsatli2018_MILPintegratedEHS,Welder2019_OptimizationofEHS}, whose solution quality is arbitrary as their feasible region does not contain the original feasible region defined by Problem~\ref{OptimalIntegratedPlanning}.}


\section{Numerical evaluation}\label{sec_numericaleval} 

The integrated modelling in this work adopts a half-hourly resolution $\Delta t = \SI{0.5}{\hour}$, which matches the granularity of the RES forecasts from the Australian Energy Market Operator (AEMO) \cite{AustralianEnergyMarketOperator2022_ISPDatabase}. The lifespan of the project is assumed to be 20 years and the discount factor is assumed to be 6\% ($\iota = 0.06$). However, instead of also considering a planning horizon of $\left| \mathcal{Y} \right| = 20$ years, which would translate to millions of variables and constraints that would make Problem~\ref{OptimalIntegratedPlanning_MIQCP} intractable, only one representative year is carefully chosen in the planning horizon, i.e., $\left| \mathcal{Y} \right| = 1$. In a similar attempt to reduce the size of the problem while still capturing the necessary variability in RES, only 4 representative weeks, one in each season, are chosen in this representative year. This translates to $ (24/\Delta t) \times 7 \times 4 = 1344$ time steps, i.e., $t \in \mathcal{T}=\{1,2,\ldots, 1344 \}$, which still results in large-scale problems as will be shown in the case studies below. As a result, the hydrogen sales in the subsequent 19 years are assumed to be constant cash flows. The corresponding term in the objective function therefore becomes
\begin{align*}
	C^{\rm H_2} = \sum_{t \in \mathcal{T}} \sum_{m \in \mathcal{J}} \frac{13.04 c_{m}^{\rm H_2} \phi_{mt}^{\rm H_2,d} \Delta \tau}{\iota}\left( 1 - \frac{1}{(1+\iota)^{20}}\right) 
\end{align*}
which is effectively the net present value of the hydrogen sales. Since 28 representative days are considered in the representative year, the selling price is multiplied by $365.25/28=13.04$ to cover a whole year's worth of H$_2$ sales. The hydrogen selling price is taken from \cite{ARENA2018_H2Exports} as \SI{3.23}{\$\per\kilogram}, i.e., $c_{m}^{\rm H_2} = 3.23 \rho = \SI{0.277}{\$\per\meter\cubed}$. The minimum and maximum operating pressures in the H$_2$ network are assumed to be \SI{3.5}{\mega\pascal} and \SI{10}{\mega\pascal}, respectively.\footnote{All the costs in this paper are in US dollars.} The cost and parameter assumptions of different transport technologies are shown in Tables~\ref{table_PtGassumptions}-\ref{table_HVACassumptions}.
\begin{table}[t!]
	\centering
	\begin{tabular}{l c} 
		\hline\hline
		\multicolumn{2}{c}{\textbf{Electrolyser} \cite{Reuß2017_SeasonalStorageH2supplychain,Reuß2019_H2SupplyChainwithSpatialResolution}} \\
		\hline
		Cost (\SI{}{\mega\$\per\mega\watt}) & 0.6 \\ 
		Efficiency $\eta_{mn}^{\rm ptg}$ (\%) & 70 \\
		Water consumption (\SI{}{\kilogram\per\kilogram}H$_2$ ) & 10 \\
		\hline
		\multicolumn{2}{c}{\textbf{Compressor} \cite{Jens2021_ExtendedEuropeanH2backbone}} \\
		\hline
		Cost (\SI{}{\mega\$\per\mega\watt}) & 4.15 \\
		Inlet pressure (\SI{}{\mega\pascal}) & 3.5 \\
		Outlet pressure (\SI{}{\mega\pascal}) & 10 \\
		\textcolor{black}{Efficiency $\eta_{im}^{\rm cp}$ (\%) } & \textcolor{black}{81} \\
		\hline
	\end{tabular}
	\caption{Cost and parameter assumptions of electrolysers \cite{Reuß2017_SeasonalStorageH2supplychain,Reuß2019_H2SupplyChainwithSpatialResolution} and compressors \cite{Jens2021_ExtendedEuropeanH2backbone} in electrolyser stations (see Figure~\ref{fig_ptg}).}
	\label{table_PtGassumptions}
\end{table}
\begin{table}[t!]
	\centering
	\begin{tabular}{l c c c} 
		\hline\hline 
		\multicolumn{4}{c}{\textbf{H$_2$ pipelines}} \\
		\hline
		\textbf{Diameter} (\SI{}{\meter}) & \textbf{0.5} & \textbf{0.9} & \textbf{1.2} \\
		\hline
		Cost (\SI{}{\mega\$\per\km}) & 1.829 & 2.682 & 3.414 \\ 
		Efficiency $\eta_{mn}^{o}$ (\%) & 95 & 95 & 95 \\
		Minimum pressure (\SI{}{\mega\pascal}) & 3.5 & 3.5 & 3.5 \\
		Maximum pressure (\SI{}{\mega\pascal}) & 10 & 10 & 10 \\
		\hline
	\end{tabular}
	\caption{Cost and parameter assumptions of H$_2$ pipelines \cite{Jens2021_ExtendedEuropeanH2backbone}.}
	\label{table_pipelineassumptions}
\end{table}
\begin{table}[t!]
	\centering
	\begin{tabular}{l c c c} 
		\hline\hline 
		\multicolumn{4}{c}{\textbf{VSC HVDC \SI{500}{\kilo\volt}}} \\
		\hline
		\textbf{Capacity} (\SI{}{\giga\watt}) & \textbf{1} & \textbf{2} & \textbf{3} \\
		\hline
		Conductor cost (\SI{}{\mega\$\per\km}) & 0.78 & 0.95 & 1 \\ 
		Converter stations (\SI{}{\mega\$}) & 2$\times$170 & 2$\times$237.4 & 2$\times$301.4 \\
		Conductor resistance (\SI{}{\ohm}) & 0.0059 & 0.0059 & 0.0059 \\
		$\alpha$ (\SI{}{\mega\watt}) & 6.62 & 6.62 & 6.62 \\
		$\beta$ (\SI{}{\volt}) & 1800 & 1800 & 1800 \\
		$\gamma$ (\SI{}{\ohm}) & 1.98 & 1.98 & 1.98 \\
		\hline
	\end{tabular}
	\caption[]{Cost and parameter assumptions of VSC HVDC systems \cite{Pletka2014_CostofTransmissionandSubstations,Bahrman2007_ABCsofHVDC,Daelemans2009_MinLossesusingHVDCVSC}.\footnotemark}
	\label{table_HVDCassumptions}
\end{table}
\begin{table}[t!]
	\centering
	\resizebox{\linewidth}{!}{
	\begin{tabular}{l | l c c c} 
		\hline\hline 
		\multicolumn{2}{c}{\textbf{Voltage (\SI{}{\kilo\volt})}} & \textbf{365} & \textbf{500} & \textbf{765} \\
		\hline
		Conductor cost & Single circuit & 0.84 & 1.19 & 1.67 \\ 
		\cline{2-5}
		(\SI{}{\mega\$\per\km}) & Double circuit & 1.34 & 1.91 & 2.38 \\ 
		\hline
		Substation cost & Single circuit & 2$\times$0.0706 & 2$\times$0.0986 & 2$\times$0.1168 \\ 
		\cline{2-5}
		(\SI{}{\mega\$\per\km}) & Double circuit & 2$\times$0.1412 & 2$\times$0.1971 & 2$\times$0.2336 \\ 
		\hline
		Capacity & Single circuit & 750 & 1500 & 1500 \\ 
		\cline{2-5}
		(\SI{}{\mega\watt}) & Double circuit & 1500 & 3000 & 3000 \\ 
		\hline
		\multicolumn{2}{l}{$r_{ij}^{o}$ (\SI{}{\ohm\per\kilo\meter})} & 0.0339 & 0.0226 & 0.01695 \\ \hline
		\multicolumn{2}{l}{$x_{ij}^{o}$ (\SI{}{\ohm\per\kilo\meter})} & 0.288 & 0.276 & 0.278 \\ \hline
		\multicolumn{2}{l}{$b_{ij}^{{\rm ch},o}$ (\SI{}{\micro\siemens\per\kilo\meter})} & 3.803 & 3.968 & 3.937 \\ \hline
		\multicolumn{2}{l}{\textcolor{black}{Voltage range}} & $\pm \SI{10}{\%}$ & $\pm \SI{10}{\%}$ & $\pm \SI{10}{\%}$ \\ \hline
		\multicolumn{2}{l}{Maximum angular displacement} & \SI{45}{\degree} & \SI{45}{\degree} & \SI{45}{\degree} \\ \hline
		\multicolumn{2}{l}{SVC (\SI{}{\$\per\mega\VAr})} & 88,000 & 88,000 & 88,000 \\ \hline
	\end{tabular}}
	\caption{Cost and parameter assumptions of HVAC systems \cite{Pletka2014_CostofTransmissionandSubstations,Bahrman2007_ABCsofHVDC,Glover2011_PowerSystemAnalysis,PrysmianGroup2015}.}
	\label{table_HVACassumptions}
\end{table}
\footnotetext{At the time of writing this paper, the largest commissioned \SI{500}{\kilo\volt} VSC HVDC project has a capacity of \SI{700}{\mega\watt}. However, \SI{1}{\giga\watt}, \SI{2}{\giga\watt}, and \SI{3}{\giga\watt} \SI{500}{\kilo\volt} VSC HVDC technology is assumed to be available in the near future. The costs for those are assumed to be the same as existing LCC HVDC technology of the same capacity.}

Since the base installation costs of electrolyser stations and SVCs are negligible compared to total station cost, both $c_{i,0}^{\rm ptg}$ and $c_{i,0}^{\rm var}$ are assumed to be zero, which obviates the need for binary variables, i.e., $z_{im}^{\rm ptg}$ and $z_{i}^{\rm var}$ are no longer needed. It should be emphasised that the cost and parameter assumptions in Tables~\ref{table_PtGassumptions}-\ref{table_HVACassumptions} are for the sole purpose of demonstrating the novel integrated modelling in Problem~\ref{OptimalIntegratedPlanning_MIQCP}. Therefore, in the context of these specific costs and parameter assumptions, the findings in this paper should be considered solely for illustration and demonstration purposes rather than real guidelines for energy infrastructure planners and stakeholders, for which specific studies based on agreed input data and assumptions should be performed.

In this implementation setup, \textsc{Julia} v1.7.2 \cite{Bezanson2017_Julia} is used as a programming language along with \textsc{JuMP} v0.22.3 \cite{DunningHuchetteLubin2017_JuMP} as a mathematical modelling layer for all the optimisation problems. All simulations are conducted on a computing platform with an Intel Core i7-6820HK CPU at 2.7GHz, 64-bit operating system, and 32GB RAM. The MIQCP problem in \eqref{OptimalIntegratedPlanning_MIQCP} is solved using \textsc{Gurobi} v9.5.0 \cite{Gurobi2019} with the branch-and-bound algorithm which solves continuous QCP relaxations at each node. In contrast, the linearised outer-approximation approach performed poorly on this problem. 

The proposed MIQCP formulation in Problem~\ref{OptimalIntegratedPlanning_MIQCP} is demonstrated on two case studies, one consisting of a canonical two-node system, and one involving actual renewable energy zones (REZ) in Queensland, Australia. In both case studies the profiles of RES forecasts are obtained from AEMO's Integrated System Plan (ISP) ``Central'' scenario \cite{AustralianEnergyMarketOperator2022_ISPDatabase}.

\subsection{Case study 1: Canonical 2-node system}\label{sec_casestudy} 

\begin{figure}[t!p]
	\centering
	\begin{subfigure}{1\columnwidth}
		\includegraphics[width=\columnwidth]{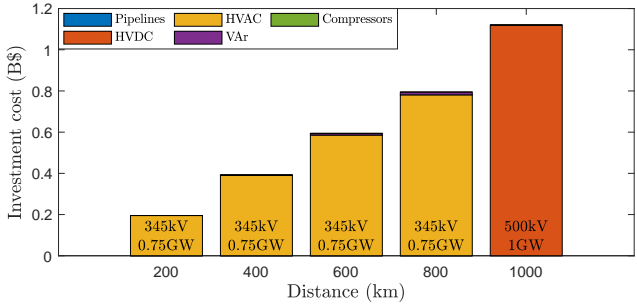}
		\caption{\SI{500}{\mega\watt} wind share and \SI{500}{\mega\watt} solar share.}
		\label{fig_2node_1GW}
	\end{subfigure}    
	\begin{subfigure}{1\columnwidth}
		\includegraphics[width=\columnwidth]{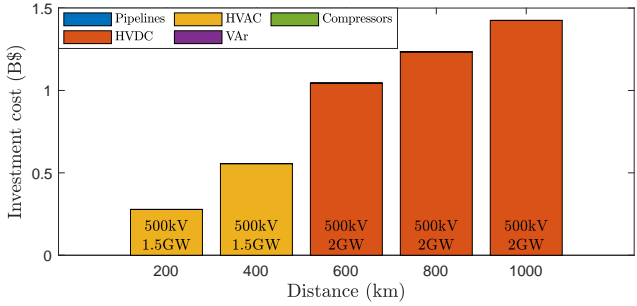}
		\caption{\SI{1250}{\mega\watt} wind share and \SI{1250}{\mega\watt} solar share.}
		\label{fig_2node_2p5GW}
	\end{subfigure}    
	\begin{subfigure}{1\columnwidth}
		\includegraphics[width=\columnwidth]{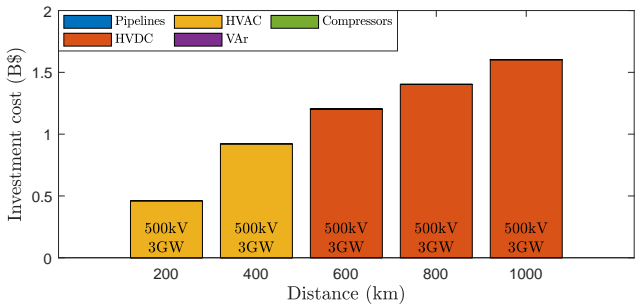}
		\caption{\SI{2500}{\mega\watt} wind share and \SI{2500}{\mega\watt} solar share.}
		\label{fig_2node_5GW}
	\end{subfigure}  
	\begin{subfigure}{1\columnwidth}
		\includegraphics[width=\columnwidth]{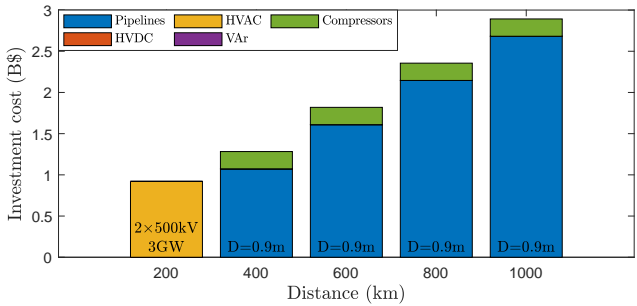}
		\caption{\SI{3750}{\mega\watt} wind share and \SI{3750}{\mega\watt} solar share.}
		\label{fig_2node_7p5GW}
	\end{subfigure}
	\begin{subfigure}{1\columnwidth}
		\includegraphics[width=\columnwidth]{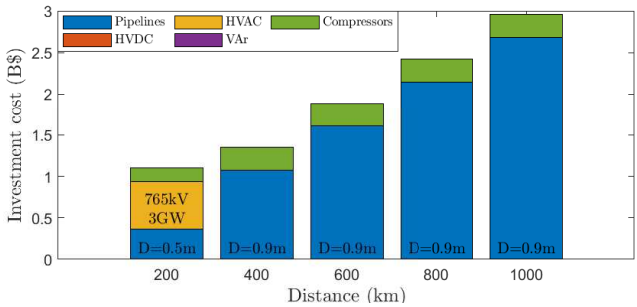}
		\caption{\SI{5000}{\mega\watt} wind share and \SI{5000}{\mega\watt} solar share.}
		\label{fig_2node_10GW}
	\end{subfigure}           
	\caption{Optimal infrastructure investment decision for different distances and RES capacity shares for the canonical 2-node system. Note the difference in y-axis scale between the figures.}
	\label{fig_2nodesystem}
\end{figure}

This case study is intended to thoroughly analyse the two fundamental drivers, namely distance and RES capacity, affecting the investment decision between two nodes. In particular, the RES capacity is varied from \SI{1}{\giga\watt} and \SI{10}{\giga\watt} and the distance is varied from \SI{200}{\kilo\meter} to \SI{1000}{\kilo\meter}, and the results are shown in Figure~\ref{fig_2nodesystem}. The RES capacity is shared equally between wind and solar and a single solar profile and a single wind profile are used, with capacity factors of 0.2731 and 0.4041, respectively. The immediate inference that can be drawn from Figure~\ref{fig_2nodesystem} is that H$_2$ pipelines tend to be preferred for higher RES capacities (\SI{7500}{\mega\watt} and above) transmitted over medium to long distances (\SI{400}{\kilo\meter} and above), whereas lower capacities (\SI{5000}{\mega\watt} and below) are dominated by electricity options. In particular, HVAC systems are the preferred option for short distances (\SI{200}{\kilo\meter} and below) across all RES capacities and HVDC systems are the preferred option for medium to long distances (\SI{600}{\kilo\meter} and above) and medium RES capacities (\SI{2500}{\mega\watt} and \SI{5000}{\mega\watt}).

Despite lower losses in HVDC systems, the high cost of converter stations places them at a disadvantage compared to HVAC systems for short to medium distances and an RES capacity of \SI{2500}{\mega\watt} to \SI{5000}{\mega\watt}. However, the larger cost of overhead conductors of HVAC tips the scale in favour of HVDC systems for medium to long distances. Additionally, angle displacement constraints on HVAC systems require SVCs to absorb the large reactive power induced by inductive and capacitive effects of long distance AC transmission, which further increases the cost of HVAC links. These results are congruent with HVAC vs HVDC comparisons in existing literature, which identify a break-even distance of around \SI{600}{\kilo\meter}, beyond which HVDC becomes more competitive \cite{DeSantis2021_Comparisonoftransportoptions}. At smaller RES capacities (\SI{1000}{\mega\watt}), where the capacity factors of 0.2731 and 0.4041 for solar and wind, respectively, translate to an average available RES power of \SI{338}{\mega\watt}, a \SI{345}{\kilo\volt} \SI{750}{\mega\watt} HVAC with an SVC at node 1 is more cost-effective than a \SI{1}{\giga\watt} \SI{500}{\kilo\volt} VSC HVDC system for transmission distances below \SI{800}{\kilo\meter}.

On the other hand, the high operating pressure range (\SI{3.5}{\mega\pascal} to \SI{10}{\mega\pascal}) of H$_2$ pipelines translates to much smaller transmission losses and large linepack capacities that together give H$_2$ pipelines an edge over electricity options for higher RES capacities (\SI{7500}{\mega\watt} and above) transmitted over medium to long distances (\SI{400}{\kilo\meter} and above). The linepack can be thought of as a large storage element that can smooth out the variability of RES.

Each one of the MIQCP problems in Figure~\ref{fig_2nodesystem} has more than 400,000 continuous variables, 28 binary variables, and more than 900,000 constraints including more than 69,000 (convex) quadratic constraints. It takes Gurobi between 1 and 3 hours to solve each one.

\subsection{Case study 2: REZ in Queensland, Australia}\label{sec_casestudy2} 

\begin{figure}[t!]
	\centering{
		\includegraphics[width=0.92\columnwidth]{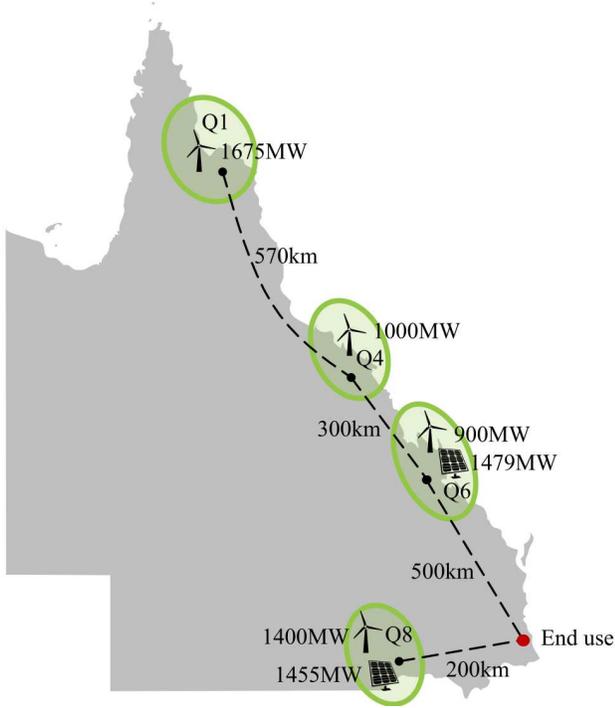}}
	\caption{RES capacity in 2040 according to AEMO's REZ \cite{AustralianEnergyMarketOperator2020_REZ}.}
	\label{fig_REZ}
\end{figure}

\begin{figure}[t!]
	\centering{
		\includegraphics[width=1\columnwidth]{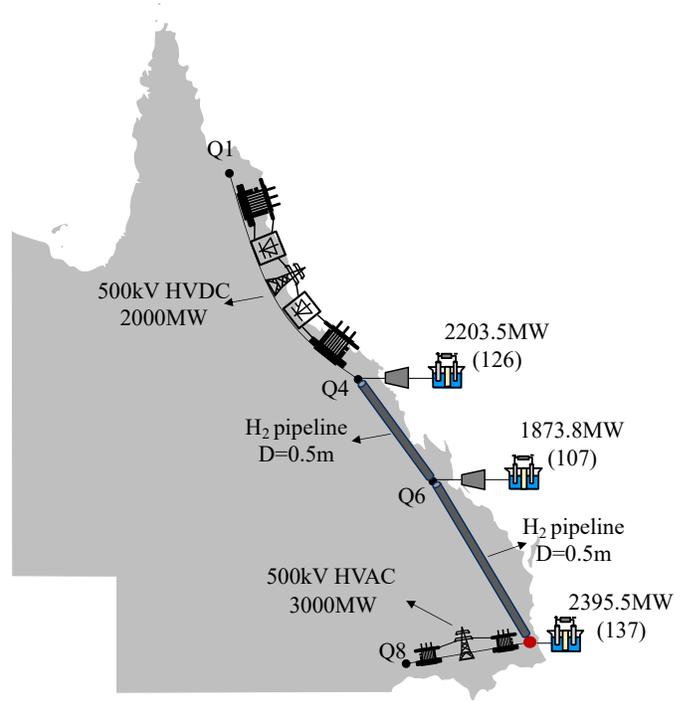}}
	\caption{Optimal solution to the infrastructure planning problem in Figure~\ref{fig_REZ}.}
	\label{fig_REZsolution}
\end{figure}

AEMO's ISP identifies potential renewable energy zones (REZ) across the national electricity market (NEM) \cite{AustralianEnergyMarketOperator2020_REZ}. In more detail, this case study considers four REZ in Queensland and one H$_2$ demand point (off-take), as shown in Figure~\ref{fig_REZ}. Figure~\ref{fig_REZ} also shows the RES capacity forecast for 2040. The solution to this integrated planning problem, shown in Figure~\ref{fig_REZsolution}, is a hybrid system consisting of a \SI{2}{\giga\watt} VSC HVDC link between REZ Q1 and Q4 (\SI{570}{\kilo\meter}), a \SI{3}{\giga\watt} \SI{500}{\kilo\volt} double circuit HVAC link between REZ Q8 and the demand point (\SI{200}{\kilo\meter}), and a \SI{0.5}{\meter} diameter H$_2$ pipeline between REZ Q4 and Q6 (\SI{300}{\kilo\meter}) and Q6 and the demand point (\SI{500}{\kilo\meter}). No VAr plants (SVCs) were needed in this case as the HVAC link is installed over the relatively short distance of \SI{200}{\kilo\meter} where both the voltage drop and angle difference (and also power losses) are small. These results are in congruence with the 2-node results in the previous section.

\begin{figure*}[t!]
	\centering{
		\psfrag{GW}{\footnotesize $\SI{}{\giga\watt}$ \normalsize}
		\psfrag{Time}{\footnotesize Time $(\mathcal{T})$ \normalsize}
		\psfrag{Available VRE}{\footnotesize Available VRE ($\overline{p}^{\rm res}_{ity}$) \normalsize}
		\psfrag{Accommodated VRE}{\footnotesize Accommodated VRE ($p^{\rm res}_{ity}$) \normalsize}
		\psfrag{Demand}{\footnotesize Demand ($\phi_{mty}^{\rm H_2,d} HHV$) \normalsize}
		\includegraphics[width=1\textwidth]{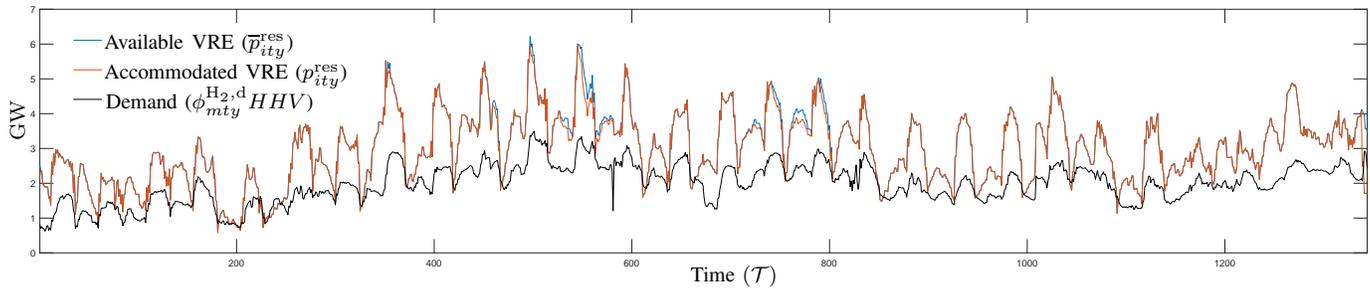}}
	\caption{Profiles of available (forecast) VRE ($\overline{p}^{\rm res}_{ity}$), accommodated VRE ($p^{\rm res}_{ity}$), and demand ($\phi_{mty}^{\rm H_2,d} HHV$) at the optimal solution of Problem~\ref{OptimalIntegratedPlanning_MIQCP}.}
	\label{fig_RESandDemand}
\end{figure*}

\begin{figure*}[t!]
	\centering{
		\psfrag{TJ}{\footnotesize $\SI{}{\tera\joule}$ \normalsize}
		\psfrag{Time}{\footnotesize Time $(\mathcal{T})$ \normalsize}
		\includegraphics[width=1\textwidth]{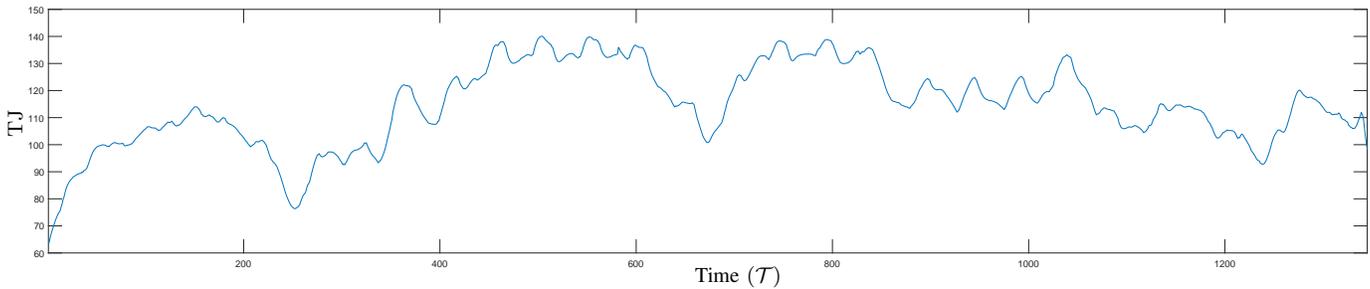}}
	\caption{Linepack profile in the two H$_2$ pipelines installed between REZ Q4 and Q6 (\SI{300}{\kilo\meter}) and Q6 and the demand point (\SI{500}{\kilo\meter}).}
	\label{fig_linepack}
\end{figure*}

Finally, the profiles of available (forecast) VRE, accommodated VRE ($p^{\rm res}_{ity}$), and demand ($\phi_{mty}^{\rm H_2,d} HHV$) at the optimal solution of Problem~\ref{OptimalIntegratedPlanning_MIQCP} are shown in Figure~\ref{fig_RESandDemand}. It can be seen from Figure~\ref{fig_RESandDemand} that the demand profile varies over a smaller range compared to the profile of available (forecast) VRE input and this is due to the effect of the linepack in the two H$_2$ pipelines installed between REZ Q4 and Q6 (\SI{300}{\kilo\meter}) and Q6 and the demand point (\SI{500}{\kilo\meter}). This linepack profile is shown in Figure~\ref{fig_linepack}. Figure~\ref{fig_RESandDemand} also shows that the optimal infrastructure investment planning in Figure~\ref{fig_REZsolution} has an \emph{energy transmission factor} of 0.9901, which means that 99.01\% of the total generated VRE is accommodated by the installed infrastructure. Recall that the energy transmission factor is defined as 
\begin{align}
	TF = \frac{ \displaystyle \underset{t \in \mathcal{T}}{\sum} \underset{y \in \mathcal{Y}}{\sum} \underset{i \in \mathcal{B}}{\sum} p^{\rm res}_{ity}}{ \displaystyle \underset{t \in \mathcal{T}}{\sum} \underset{y \in \mathcal{Y}}{\sum} \underset{i \in \mathcal{B}}{\sum} \overline{p}^{\rm res}_{ity}} . 
\end{align}


The MIQCP problem in this case study has around than 560,000 continuous variables, 44 binary variables, and more than 1,160,000 constraints including more than 107,000 (convex) quadratic constraints. It takes Gurobi around 4 days to solve it.

\section{Conclusion}\label{sec_conclusion}

To address the challenging question of whether to transport large-scale VRE as molecules in H$_2$ pipelines or as electrons in electricity transmission lines, this paper introduced a first-of-its-kind mathematical framework for finding the optimal integrated planning of electricity and H$_2$ infrastructure. The model fills the gap in existing state-of-the-art literature by (i) considering all relevant infrastructure technologies such as HVDC, HVAC, SVCs, and H$_2$ pipelines and compressors, and by (ii) incorporating essential nonlinearities such as voltage drops due to losses in HVAC and HVDC transmission lines, losses in HVDC converter stations, reactive power flow, pressure drops in pipelines, and linepack, all of which play an important role in determining the optimal infrastructure investment decision. The high temporal resolution of the RES forecasts makes this model a large-scale nonconvex MINLP problem that is intractable if solved directly using MINLP solvers. The paper therefore proposes a tractable alternative in the form of an MIQCP relaxation that is demonstrated on a canonical two-node system as well as on a real-world case study involving actual renewable energy zones in Australia. 
	
\section*{Acknowledgment}

This work is supported by Future Fuels Cooperative Research Centre as part of the RP1.1-02B: ``Transport and Storage Options for Future Fuels'' project. The cash and in-kind support from the industry participants is gratefully acknowledged.

\printbibliography

\end{document}